\documentclass[%
preprint,
 amsmath,amssymb,notitlepage,
 aps,showkeys
]{revtex4-2}

\usepackage{graphicx}
\usepackage{dcolumn}
\usepackage{bm}

\usepackage{graphicx}             
\usepackage{array}
\usepackage{xmpmulti}
\usepackage{amsmath,latexsym,amssymb,fixmath,tipa}
\usepackage{mathrsfs}
\usepackage{float}
\usepackage{rotating}
\usepackage{xcolor}
\usepackage{soul}
\usepackage{transparent}
\usepackage{ragged2e}
\usepackage{bigints}

\usepackage{xcolor}
\usepackage{tcolorbox}

\usepackage{pgfplots}
\pgfplotsset{compat=newest}
\usepackage{caption}
\usepackage[list=true]{subcaption}

\usepackage[compact]{titlesec}
\titlespacing{\section}{0pt}{0ex}{0ex}
\titlespacing{\subsection}{0pt}{-2ex}{0ex}
\titlespacing{\subsubsection}{0pt}{-2ex}{0ex}

\usepackage{caption,setspace}
\usepackage{fancyhdr}
\usepackage{datetime}
\def\mixedindices#1#2{\mathstrut_{#1}^{#2}}

\def\u#1#2{u \mixedindices{#1}{#2}}

\def\ubar#1#2{\overset{*}{u} \mixedindices{#1}{#2}}

\def\Umatbar#1#2{\overset{*}{U} \mixedindices{#1}{#2}}

\def\a#1#2{a \mixedindices{#1}{#2}}
\def\abar#1#2{\overset{*}{a} \mixedindices{#1}{#2}}
\def\b#1#2{b \mixedindices{#1}{#2}}
\def\bbar#1#2{\overset{*}{b} \mixedindices{#1}{#2}}
\def\c#1#2{c \mixedindices{#1}{#2}}
\def\cbar#1#2{\overset{*}{c} \mixedindices{#1}{#2}}
\def\d#1#2{d \mixedindices{#1}{#2}}
\def\dbar#1#2{\overset{*}{d} \mixedindices{#1}{#2}}

\def\Gmatbar#1#2{\overset{*}{G} \mixedindices{#1}{#2}}
\def\Gmat#1#2{G \mixedindices{#1}{#2}}

\def\s#1#2{s \mixedindices{#1}{#2}}
\def\sbar#1#2{\overset{*}{s} \mixedindices{#1}{#2}}

\def\vvec#1#2{ {\rm v} \mixedindices{#1}{#2}}
\def\vvechat#1#2{ \hat{{\rm v}} \mixedindices{#1}{#2}}
\def\wvec#1#2{ {\rm w} \mixedindices{#1}{#2}}
\def\wvechat#1#2{ \hat{{\rm w}} \mixedindices{#1}{#2}}

\def\lam#1#2{\lambda \mixedindices{#1}{#2}}
\def\lamhat#1#2{\hat{\lambda} \mixedindices{#1}{#2}}
\def\lambar#1#2{\overset{*}{\lambda} \mixedindices{#1}{#2}}
\def\lambarhat#1#2{\overset{*}{\hat{\lambda}} \mixedindices{#1}{#2}}

\def\sbar#1#2{\overset{*}{s} \mixedindices{#1}{#2}}

\def\dinv{\partial^{-1}}

\def\i{{\rm i}}
\def\imag{{\rm Im}}
\def\real{{\rm Re}}

\def\Pbar#1#2{\overset{*}{P} \mixedindices{#1}{#2}}
\def\Wbar#1#2{\overset{*}{W} \mixedindices{#1}{#2}}
\def\Vbar#1#2{\overset{*}{V} \mixedindices{#1}{#2}}
\def\Sbar#1#2{\overset{*}{S} \mixedindices{#1}{#2}}
\def\phibar#1#2{\overset{*}{\phi}\mixedindices{#1}{#2}}

\def\Dsol#1#2{\mathsf{D} \mixedindices{#1}{#2}}
\def\Nsol#1#2{\mathsf{N} \mixedindices{#1}{#2}}
\def\Gsol#1#2{\mathsf{G} \mixedindices{#1}{#2}}
\def\Gbarsol#1#2{\overset{*}{\mathsf{G}} \mixedindices{#1}{#2}}

\def\Gammaop#1#2{\Gamma \mixedindices{#1}{#2}}
\def\Gammabarop#1#2{\overset{*}{\Gamma} \mixedindices{#1}{#2}}

\def\Identityop{{\rm Id}}

\usepackage{amsthm}

\newtheorem{remark}{Remark}

\begin{document}


\title{Riemann-Hilbert problems for a nonlocal reverse-spacetime Sasa-Satsuma hierarchy of a fifth-order equation and its soliton solutions}

\author{Ahmed M. G. Ahmed$^{1,2,}$}
\email[Corresponding author:]{ahmedmgahmed@usf.edu}
\author{Alle Adjiri$^{1,}$}
\email{aadjiri@usf.edu}
\author{Solomon Manukure $^{3,}$}
\email{solomon.manukure@famu.edu}
\affiliation{$^{1}$Department of Mathematics and Statistics
\vspace{-1mm}
\\ 
University of South Florida
\vspace{-1mm}
\\ 
Tampa, FL 33620-5700,
USA}
\affiliation{$^{2}$Department of Mathematics and Statistics
\vspace{-1mm}
\\
Brock University
\vspace{-1mm}
\\
St.Catharines, ON L2S3A1, Canada}
\affiliation{$^{3}$Department of Mathematics 
\vspace{-1mm}
\\ 
Florida A $\&$ M University
\vspace{-1mm}
\\ 
Tallahassee, FL 32307,
USA}
\date{\today}
\begin{abstract}
We aim to present and analyze a nonlinear nonlocal
reverse-spacetime  fifth-order scalar Sasa-Satsuma equation, based on a nonlocal $5 \times 5$ matrix AKNS spectral problem. Starting from a nonlocal matrix AKNS spectral problem, local and nonlocal symmetry relations are derived from a group of rotations. A kind of Riemann-Hilbert problem is formulated,
which allows to generate soliton solutions by using vectors lying in the kernel of the matrix Jost solutions. When reflection coefficients are zeros, the jump matrix is the identity and the corresponding Riemann-Hilbert problem yields soliton solutions, whose explicit formulas enable us to explore their dynamics.
\end{abstract}

\maketitle
\newpage
\section{Introduction}
The study of integrable systems has been one of the most fascinating branches of mathematics, and an intriguing area for both mathematicians and physicists. Integrable systems and their properties can be used to predict many natural phenomena. They are commonly found in nonlinear optics, plasmas, ocean and water waves, gravitational fields, and fluid dynamics \cite{KangXia2019}-\cite{Osborne2010}.
The Korteweg-de Vries (KdV) equation, the nonlinear Schr\"{o}dinger (NLS) equation and the Kadomtsev-Petviashvili (KP) equation are typical examples of integrable systems. Integrable systems can provide a variety of soliton solutions, for example breathers, lumps, and rogue waves.
Soliton solutions are a kind of special solutions which are stable 
localized waves. Nonlocal PT symmetric reverse-spacetime, reverse-time,
and reverse-space have been studied for the NLS and KdV equations under 
the inverse scattering transformation and the Riemann-Hilbert problem
\cite{AblowitzMusslimani2016}-\cite{LingMa}. 
Riemann-Hilbert formulation is an interesting method to investigate
and explore new examples of nonlocal integrable equations and their soliton solutions \cite{Yang2018}-\cite{Ma2021}. 
In this paper, we present a new nonlocal reverse-spacetime
fifth-order Sasa-Satsuma equation \cite{Geng1}-\cite{Geng2}:
\begin{flalign}\label{Sasa-Satsuma-equation-u-SASA}
\u{t}{}&=\u{xxxxx}{}
-10 (|u|^{2}+|u(-x,-t)|^{2}) \u{xxx}{}
-15(|u|^{2}+|u(-x,-t)|^{2})_{x} \u{xx}{}
\\ \nonumber
&
+ \big[ -15 (|u|^{2}+|u(-x,-t)|^{2})_{xx}
+10 (|\u{x}{}|^{2}+|\u{x}{}(-x,-t)|^{2})
+ 40 (|u|^{2}+|u(-x,-t)|^{2})^{2} \big] \u{x}{}
\\& \nonumber
+ \big[
-5 (|u|^{2}+|\u{}{}(-x,-t)|^{2})_{xxx}
+5 (|\u{x}{}|^{2}+|\u{x}{}(-x,-t)|^{2})_{x}
+20 \big( (|u|^{2}+|\u{}{}(-x,-t)|^{2})^{2} \big)_{x}
\big] \u{}{}.
\end{flalign}
We formulate a kind of Riemann-Hilbert problems for the above
integrable nonlocal Sasa-Satsuma equation with the real line
being its contour, and solve the resulting Riemann-Hilbert problems with identity jump matrix to present its soliton solutions \cite{KangXiaMa2019}-\cite{MaActa2022}.
\newline
The outline of the paper is as follows.
In section 2, we construct a nonlocal Sasa-Satsuma hierarchy associated with a nonlocal $5 \times 5$ matrix AKNS spectral problem and its Hamiltonian structure. In section 3, we formulate a kind of Riemann-Hilbert problems, based on the corresponding matrix spectral problems. In
section 4, we compute soliton solutions by taking the reflection coefficients to be zeros \cite{Yang2010}-\cite{DrazinJohnson1989}.
In section 5, we present explicit and exact one soliton solution, and classify the different cases for the explicit two soliton solutions, and explore their dynamical behaviours. The last section will be a brief conclusion and some remarks.
\section{AKNS hierarchy}\noindent 
Consider the nonlocal $5 \times 5$ matrix AKNS spatial spectral problem
\cite{Ma2018}
\begin{align}\label{Spatialspectralproblem-SASA}
\psi_{x} &= \i U \psi,
\end{align}
where $\psi$ is the eigenfunction and the spectral matrix $U(\u{}{},\lambda)$ is given by
\begin{equation}\label{Umatrix-SASA}
U(u,\lambda)=\begin{pmatrix}
\alpha_{1} \lambda & \u{}{}(x,t) & \u{}{}(-x,-t) & 
\ubar{}{}(x,t) & \ubar{}{}(-x,-t)
\\
-\ubar{}{}(x,t) & \alpha_{2} \lambda & 0 & 0 & 0
\\
-\ubar{}{}(-x,-t) & 0 & \alpha_{2} \lambda & 0 & 0
\\
-\u{}{}(x,t) & 0 & 0 & \alpha_{2} \lambda & 0
\\
-\u{}{}(-x,-t) & 0 & 0 & 0 & \alpha_{2} \lambda
\end{pmatrix}
=\lambda \Lambda + P(u),
\end{equation}
where $\Lambda=diag(\alpha_{1},\alpha_{2} I_{4})$, $\lambda$ is a nonzero spectral parameter, $\alpha_{1},\alpha_{2}$
are two distinct real constants, and 
$u$ is the potential. We assume that $\u{}{}$ and 
$x \u{}{}$ belong to the $L^{2}$ space and
\begin{equation}\label{Pmatrix-SASA}
P=\begin{pmatrix}
0 & \u{}{}(x,t) & \u{}{}(-x,-t) & \ubar{}{}(x,t) & \ubar{}{}(-x,-t)
\\
-\ubar{}{}(x,t) & 0 & 0 & 0 & 0
\\
-\ubar{}{}(-x,-t) & 0 & 0 & 0 & 0
\\
-\u{}{}(x,t) & 0 & 0 & 0 & 0
\\
-\u{}{}(-x,-t) & 0 & 0 & 0 & 0
\end{pmatrix}.
\end{equation}
\begin{remark}
One can see that the matrix $U$ has the
following simultaneous symmetry relations:
\begin{align}\label{Ureduction-SASA}
\begin{cases}
U^{\dagger}(x,t,-\lambar{}{})=-C_{0}U(x,t,\lam{}{}) C_{0}^{-1}
=-U(x,t,\lam{}{}), 
\quad
U(x,t,-\lam{}{})=-C_{4}U(x,t,\lam{}{}) C_{4}^{-1}
\\ 
U^{T}(x,t,-\lam{}{})=-C_{1}U(x,t,\lam{}{})C_{1}^{-1},
\quad
U^{T}(x,t,\lam{}{})=C_{5}U(x,t,\lam{}{})C_{5}^{-1},
\\
U^{\dagger}(-x,-t,-\lambar{}{})=-C_{2}U(x,t,\lambda)C_{2}^{-1},
\quad
U^{\dagger}(-x,-t,\lambar{}{})=C_{6}U(x,t,\lambda)C_{6}^{-1},
\\
\Umatbar{}{}(-x,-t,\lambar{}{})=C_{3}U(x,t,\lam{}{})C_{3}^{-1},
\quad 
\Umatbar{}{}(-x,-t,-\lambar{}{})=-C_{7}U(x,t,\lam{}{})C_{7}^{-1},
\end{cases}
\end{align}
\newpage
where the eight $5 \times 5$ matrices are
\begin{align}
C_{i}&=
\begin{pmatrix}
1 & \mathbf{0_{14}} 
\\
\mathbf{0_{41}} & \sigma_{i}
\end{pmatrix},
\quad i \in \{0,\ldots,7\},
\end{align}
with $\mathbf{0_{14}}$, $\mathbf{0_{41}}$ being the 
four-component zero row and column vectors respectively, and $\sigma_{i}$ are:
\begin{footnotesize}
\begin{align}
\label{C1C2matrices-SASA}
\sigma_{0}&=\begin{pmatrix}
1 & 0 & 0 & 0
\\
0 & 1 & 0 & 0 
\\
0 & 0 & 1 & 0 
\\
0 & 0 & 0 & 1 
\end{pmatrix},
\sigma_{1}=\begin{pmatrix}
0 & 0 & 1 & 0 
\\
0 & 0 & 0 & 1 
\\
1 & 0 & 0 & 0 
\\
0 & 1 & 0 & 0 
\end{pmatrix},
\sigma_{2}=\begin{pmatrix}
0 & 1 & 0 & 0 
\\
1 & 0 & 0 & 0 
\\
0 & 0 & 0 & 1 
\\
0 & 0 & 1 & 0 
\end{pmatrix},
\sigma_{3}=\begin{pmatrix}
0 & 0 & 0 & 1 
\\
0 & 0 & 1 & 0 
\\
0 & 1 & 0 & 0 
\\
1 & 0 & 0 & 0 
\end{pmatrix},
\end{align}
\end{footnotesize}
and  
\begin{align}
\sigma_{4}=-\sigma_{0},
\sigma_{5}=-\sigma_{1},
\sigma_{6}=-\sigma_{2},
\sigma_{7}=-\sigma_{3}.    
\end{align}
Note that all $C_{i}$ are symmetric and orthogonal matrices, i.e., $C_{i}=C_{i}^{T}$ and $C_{i}^{2}=I_{2}$, for $i \in \{0,\ldots,7\}$. In fact, they form an orthogonal group, 
$G=\{ C_{0},C_{1},C_{2},C_{3},C_{4},C_{5},C_{6},C_{7}\}$
that has two connected components. The first component $G_{1}$ of all matrices where $det(C_{i})=1$ for 
$i \in \{0,1,2,3\}$, that is; the normal subgroup
component $G_{1}=SO(5)=\{C_{0},C_{1},C_{2},C_{3}\}$.
The second component is $G_{2}=\{C_{4},C_{5},C_{6},C_{7}\}$,
where $det(C_{i})=-1$ for 
$i \in \{4,5,6,7\}$.
\newline
In this paper, we will consider all reductions generated by the matrices in the rotation group $SO(5)$.
\end{remark}
\noindent
In addition, since $U(x,t,\lambda)=\lambda \Lambda+P(x,t)$,
then we can easily prove that
\begin{align}\label{Pequ-SASA}
\begin{cases}
P^{\dagger}(x,t)=-P(x,t),
\\
P^{T}(x,t)=-C_{1}P(x,t)C_{1}^{-1},
\\
P^{\dagger}(-x,-t)=-C_{2}P(x,t)C_{2}^{-1},
\\
\Pbar{}{}(-x,-t)=C_{3}P(x,t)C_{3}^{-1}.
\end{cases}
\end{align}
Let's construct the associated two-component Sasa-Satsuma soliton hierarchy. 
\newpage
\noindent
To do so, we need to solve the stationary zero curvature equation
\begin{equation}\label{stationaryZC-SASA}
W_{x}=\i[U,W],     
\end{equation} 
for
\begin{equation}
W=\begin{pmatrix} 
a & b_{1} & b_{2} & b_{3} & b_{4}
\\
c_{1} & d_{11} & d_{12} & d_{13} & d_{14}
\\
c_{2} & d_{21} & d_{22} & d_{23} & d_{24}
\\
c_{3} & d_{31} & d_{32} & d_{33} & d_{34}
\\
c_{4} & d_{41} & d_{42} & d_{43} & d_{44}
\end{pmatrix},
\end{equation}
where $a,b_{i},c_{i},d_{ij}$ are scalar components for 
$i,j \in \{1,2,3,4\}$. From the stationary zero curvature equation, we get:
\begin{flalign}
a_{x}^{}&=\i \Big[
 \ubar{}{}(x) \b{1}{}
+ \ubar{}{}(-x) \b{2}{}
+ \u{}{}(x) \b{3}{}
+  \u{}{}(-x) \b{4}{}
\nonumber
\\&
+  \u{}{}(x) \c{1}{}
+  \u{}{}(-x) \c{2}{}
+  \ubar{}{}(x) \c{3}{}
+  \ubar{}{}(-x) \c{4}{}
\Big],
\\
\b{1,x}{}&= \i \big[ 
\alpha \lambda \b{1}{}
-  \u{}{}(x) \a{}{}
+  \u{}{}(x) \d{11}{}
+  \u{}{}(-x) \d{21}{}
+  \ubar{}{}(x) \d{31}{}
+  \ubar{}{}(-x) \d{41}{}
\big],
\\ 
\b{2,x}{}&= \i \big[ 
\alpha \lambda \b{2}{}
-  \u{}{}(-x) \a{}{}
+  \u{}{}(x) \d{12}{}
+  \u{}{}(-x) \d{22}{}
+  \ubar{}{}(x) \d{32}{}
+  \ubar{}{}(-x) \d{42}{}
\big],
\\ 
\b{3,x}{}&= \i \big[ 
\alpha \lambda \b{3}{}
-  \ubar{}{}(x) \a{}{}
+  \u{}{}(x) \d{13}{}
+  \u{}{}(-x) \d{23}{}
+  \ubar{}{}(x) \d{33}{}
+  \ubar{}{}(-x) \d{43}{}
\big],
\\ 
\b{4,x}{}&= \i \big[ 
\alpha \lambda \b{4}{}
-  \ubar{}{}(-x) \a{}{}
+  \u{}{}(x) \d{14}{}
+  \u{}{}(-x) \d{24}{}
+  \ubar{}{}(x) \d{34}{}
+  \ubar{}{}(-x) \d{44}{}
\big],
\\
\c{1,x}{}&= \i \big[ 
-\alpha \lambda \c{1}{}
-  \ubar{}{}(x) \a{}{}
+  \ubar{}{}(x) \d{11}{}
+  \ubar{}{}(-x) \d{12}{}
+  \u{}{}(x) \d{13}{}
+  \u{}{}(-x) \d{14}{}
\big],
\\ 
\c{2,x}{}&= \i \big[ 
-\alpha \lambda \c{2}{}
-  \ubar{}{}(-x) \a{}{}
+  \ubar{}{}(x) \d{21}{}
+  \ubar{}{}(-x) \d{22}{}
+  \u{}{}(x) \d{23}{}
+  \u{}{}(-x) \d{24}{}
\big],
\\ 
\c{3,x}{}&= \i \big[ 
-\alpha \lambda \c{3}{}
-  \u{}{}(x) \a{}{}
+  \ubar{}{}(x) \d{31}{}
+  \ubar{}{}(-x) \d{32}{}
+  \u{}{}(x) \d{33}{}
+  \u{}{}(-x) \d{34}{}
\big],
\\ 
\c{4,x}{}&= \i \big[ 
-\alpha \lambda \c{4}{}
-  \u{}{}(-x) \a{}{}
+  \ubar{}{}(x) \d{41}{}
+  \ubar{}{}(-x) \d{42}{}
+  \u{}{}(x) \d{43}{}
+ \u{}{}(-x) \d{44}{}  
\big], &
\end{flalign}
\begin{flalign}\label{recursivesystem6order-SASA}
\d{11,x}{}&= 
- \i \big[ 
\ubar{}{}(x) \b{1}{}
+  \u{}{}(x) \c{1}{}
\big],
& 
\d{12,x}{}&= 
-\i \big[ 
\ubar{}{}(x) \b{2}{}
+ \u{}{}(-x)  \c{1}{}
\big], 
\\
\d{21,x}{}&= 
-\i \big[ 
\ubar{}{}(-x) \b{1}{}
+  \u{}{}(x)  \c{2}{}
\big],
& 
\d{22,x}{}&= 
-\i \big[ 
 \ubar{}{}(-x) \b{2}{}
+ \u{}{}(-x) \c{2}{}
\big],
\\
\d{31,x}{}&= 
-\i \big[ 
 \u{}{}(x) \b{1}{}
+  \u{}{}(x) \c{3}{}
\big],
&
\d{32,x}{}&= 
-\i \big[ 
 \u{}{}(x) \b{2}{}
+ \u{}{}(-x)  \c{3}{}
\big],
\\
\d{41,x}{}&= 
-\i \big[ 
 \u{}{}(-x) \b{1}{}
+  \u{}{}(x) \c{4}{}
\big],
&
\d{42,x}{}&= 
-\i \big[ 
 \u{}{}(-x) \b{2}{}
+ \u{}{}(-x)  \c{4}{}
\big], &
\end{flalign}
\begin{flalign}
\d{13,x}{}&= 
-\i \big[ 
 \ubar{}{}(x) \b{3}{}
+ \ubar{}{}(x)  \c{1}{}
\big],
&
\d{14,x}{}&= 
-\i \big[ 
 \ubar{}{}(x) \b{4}{}
+ \ubar{}{}(-x) \c{1}{}
\big],
\\
\d{23,x}{}&= 
-\i \big[ 
 \ubar{}{}(-x) \b{3}{}
+ \ubar{}{}(x) \c{2}{}
\big],
&
\d{24,x}{}&= 
-\i \big[ 
 \ubar{}{}(-x) \b{4}{}
+ \ubar{}{}(-x) \c{2}{} 
\big],
\\
\d{33,x}{}&= 
-\i \big[ 
 \u{}{}(x) \b{3}{}
+ \ubar{}{}(x) \c{3}{}
\big],
&
\d{34,x}{}&= 
-\i \big[ 
 \u{}{}(x) \b{4}{}
+ \ubar{}{}(-x) \c{3}{}
\big],
\\
\d{43,x}{}&= 
-\i \big[ 
 \u{}{}(-x) \b{3}{}
+ \ubar{}{}(x) \c{4}{}
\big],
&
\d{44,x}{}&= 
-\i \big[ 
 \u{}{}(-x) \b{4}{}
+ \ubar{}{}(-x) \c{4}{}
\big], &
\end{flalign}
where $\alpha=\alpha_{1}-\alpha_{2}$. 
Now, we expand $W$ in Laurent series, with the components of
$W$ are written explicitly as:
\begin{align*}
a&= \sum\limits_{m= 0}^{\infty}a^{[m]}\lambda^{-m},
\quad
d_{ii}= \sum\limits_{m= 0}^{\infty} d_{ii}^{[m]}\lambda^{-m},
\, i \in \{1,2,3,4\},
\\
\b{i}{}&= \sum\limits_{m= 0}^{\infty} \b{i}{[m]}\lambda^{-m}, 
\,
i \in \{1,\ldots,4\},
\quad
\c{i}{}= \sum\limits_{m= 0}^{\infty} \c{i}{[m]}\lambda^{-m}, 
\,
i \in \{1,\ldots,4\},
\end{align*}
where $W$ is rewritten in the following form:
\begin{equation}\label{WLaurentexpansion-SASA}
W = \sum\limits_{m=0}^{\infty}W_{m}\lambda^{-m} 
\quad \text{with} \quad 
W_{m}=\begin{pmatrix} 
a^{[m]}  & b_{1}^{[m]} & b_{2}^{[m]} & b_{3}^{[m]} & b_{4}^{[m]} \\
c_{1}^{[m]} & d_{11}^{[m]} & d_{12}^{[m]} & d_{13}^{[m]} & d_{14}^{[m]}   \\
c_{2}^{[m]} & d_{21}^{[m]} & d_{22}^{[m]} & d_{23}^{[m]} & d_{24}^{[m]}   \\
c_{3}^{[m]} & d_{31}^{[m]} & d_{32}^{[m]} & d_{33}^{[m]} & d_{34}^{[m]}   \\
c_{4}^{[m]} & d_{41}^{[m]} & d_{42}^{[m]} & d_{43}^{[m]} & d_{44}^{[m]}   
\end{pmatrix},
\quad 
m \geq 0.
\end{equation}
As a consquence, the system (\ref{recursivesystem6order-SASA})
generates the recursive relations:
\begin{flalign}
b_{i}^{[0]}&=c_{i}^{[0]}=0, \quad i \in \{1,\ldots,4\},
\\ \nonumber
a_{x}^{[m]}&=\i \Big[
 \ubar{}{}(x) \b{1}{[m]}
+ \ubar{}{}(-x) \b{2}{[m]}
+ \u{}{}(x) \b{3}{[m]}
+  \u{}{}(-x) \b{4}{[m]}
\\ & \label{amrecursiverelation}
+  \u{}{}(x) \c{1}{[m]}
+  \u{}{}(-x) \c{2}{[m]}
+  \ubar{}{}(x) \c{3}{[m]}
+  \ubar{}{}(-x) \c{4}{[m]}
\Big], &
\end{flalign}
\begin{flalign}\label{b1recursiverelations-SASA}
\b{1}{[m+1]}&= \frac{1}{\alpha} \big[ 
-\i \b{1,x}{[m]}
+  \u{}{}(x) \a{}{[m]}
-  \u{}{}(x) \d{11}{[m]}
-  \u{}{}(-x) \d{21}{[m]}
-  \ubar{}{}(x) \d{31}{[m]}
-  \ubar{}{}(-x) \d{41}{[m]}
\big],
\\ 
\b{2}{[m+1]}&= \frac{1}{\alpha} \big[ 
-\i \b{2,x}{[m]}
+  \u{}{}(-x) \a{}{[m]}
-  \u{}{}(x) \d{12}{[m]}
-  \u{}{}(-x) \d{22}{[m]}
-  \ubar{}{}(x) \d{32}{[m]}
-  \ubar{}{}(-x) \d{42}{[m]}
\big],
\\ 
\b{3}{[m+1]}&= \frac{1}{\alpha} \big[ 
-\i \b{3,x}{[m]}
+  \ubar{}{}(x) \a{}{[m]}
-  \u{}{}(x) \d{13}{[m]}
-  \u{}{}(-x) \d{23}{[m]}
-  \ubar{}{}(x) \d{33}{[m]}
-  \ubar{}{}(-x) \d{43}{[m]}
\big],
\\ 
\b{4}{[m+1]}&= \frac{1}{\alpha} \big[ 
-\i \b{4,x}{[m]}
+  \ubar{}{}(-x) \a{}{[m]}
-  \u{}{}(x) \d{14}{[m]}
-  \u{}{}(-x) \d{24}{[m]}
-  \ubar{}{}(x) \d{34}{[m]}
-  \ubar{}{}(-x) \d{44}{[m]}
\big], &
\end{flalign}
\begin{flalign}
\c{1}{[m+1]}&= \frac{1}{\alpha} \big[ 
\i \c{1,x}{[m]}
-  \ubar{}{}(x) \a{}{[m]}
+  \ubar{}{}(x) \d{11}{[m]}
+  \ubar{}{}(-x) \d{12}{[m]}
+  \u{}{}(x) \d{13}{[m]}
+  \u{}{}(-x) \d{14}{[m]}
\big],
\\ 
\c{2}{[m+1]}&= \frac{1}{\alpha} \big[ 
\i \c{2,x}{[m]}
-  \ubar{}{}(-x) \a{}{[m]}
+  \ubar{}{}(x) \d{21}{[m]}
+  \ubar{}{}(-x) \d{22}{[m]}
+  \u{}{}(x) \d{23}{[m]}
+  \u{}{}(-x) \d{24}{[m]}
\big],
\\ 
\c{3}{[m+1]}&= \frac{1}{\alpha} \big[ 
\i \c{3,x}{[m]}
-  \u{}{}(x) \a{}{[m]}
+  \ubar{}{}(x) \d{31}{[m]}
+  \ubar{}{}(-x) \d{32}{[m]}
+  \u{}{}(x) \d{33}{[m]}
+  \u{}{}(-x) \d{34}{[m]}
\big],
\\ \label{c4recursiverelations-SASA}
\c{4}{[m+1]}&= \frac{1}{\alpha} \big[ 
\i \c{4,x}{[m]}
-  \u{}{}(-x) \a{}{[m]}
+  \ubar{}{}(x) \d{41}{[m]}
+  \ubar{}{}(-x) \d{42}{[m]}
+  \u{}{}(x) \d{43}{[m]}
+ \u{}{}(-x) \d{44}{[m]}  
\big]. &
\end{flalign}
\begin{flalign} \label{d11mrecursiverelation}
\d{11,x}{[m]}&= 
- \i \big[ 
\ubar{}{}(x) \b{1}{[m]}
+  \u{}{}(x) \c{1}{[m]}
\big],
&
\d{12,x}{[m]}&= 
-\i \big[ 
\ubar{}{}(x) \b{2}{[m]}
+ \u{}{}(-x)  \c{1}{[m]}
\big],
\\ \label{d22mrecursiverelation}
\d{21,x}{[m]}&= 
-\i \big[ 
\ubar{}{}(-x) \b{1}{[m]}
+  \u{}{}(x)  \c{2}{[m]}
\big],
&
\d{22,x}{[m]}&= 
-\i \big[ 
 \ubar{}{}(-x) \b{2}{[m]}
+ \u{}{}(-x) \c{2}{[m]}
\big],
\\
\d{31,x}{[m]}&= 
-\i \big[ 
 \u{}{}(x) \b{1}{[m]}
+  \u{}{}(x) \c{3}{[m]}
\big],
&
\d{32,x}{[m]}&= 
-\i \big[ 
 \u{}{}(x) \b{2}{[m]}
+ \u{}{}(-x)  \c{3}{[m]}
\big],
\\
\d{41,x}{[m]}&= 
-\i \big[ 
 \u{}{}(-x) \b{1}{[m]}
+  \u{}{}(x) \c{4}{[m]}
\big],
&
\d{42,x}{[m]}&= 
-\i \big[ 
 \u{}{}(-x) \b{2}{[m]}
+ \u{}{}(-x)  \c{4}{[m]}
\big], &
\end{flalign}
\begin{flalign}
\d{13,x}{[m]}&= 
-\i \big[ 
 \ubar{}{}(x) \b{3}{[m]}
+ \ubar{}{}(x)  \c{1}{[m]}
\big],
&
\d{14,x}{[m]}&= 
-\i \big[ 
 \ubar{}{}(x) \b{4}{[m]}
+ \ubar{}{}(-x) \c{1}{[m]}
\big],
\\
\d{23,x}{[m]}&= 
-\i \big[ 
 \ubar{}{}(-x) \b{3}{[m]}
+ \ubar{}{}(x) \c{2}{[m]}
\big],
&
\d{24,x}{[m]}&= 
-\i \big[ 
 \ubar{}{}(-x) \b{4}{[m]}
+ \ubar{}{}(-x) \c{2}{[m]} 
\big],
\\ \label{d33mrecursiverelation}
\d{33,x}{[m]}&= 
-\i \big[ 
 \u{}{}(x) \b{3}{[m]}
+ \ubar{}{}(x) \c{3}{[m]}
\big],
&
\d{34,x}{[m]}&= 
-\i \big[ 
 \u{}{}(x) \b{4}{[m]}
+ \ubar{}{}(-x) \c{3}{[m]}
\big],
\\ \label{d43d44recursiverelation}
\d{43,x}{[m]}&= 
-\i \big[ 
 \u{}{}(-x) \b{3}{[m]}
+ \ubar{}{}(x) \c{4}{[m]}
\big],
&
\d{44,x}{[m]}&= 
-\i \big[ 
 \u{}{}(-x) \b{4}{[m]}
+ \ubar{}{}(-x) \c{4}{[m]}
\big], &
\end{flalign} 
The first few involved functions can be worked out as follows:
\newpage
\begin{flalign}
&\begin{aligned}
\begin{cases}
a^{[0]} &= \beta_{1},
\\
a^{[1]} &=0,
\\
a^{[2]} &= 2 \frac{\beta}{\alpha^{2}} 
\big( 
|\u{}{}|^{2}
+ |\u{}{}(-x,-t)|^{2} 
\big),
\\
a^{[3]} &= 0,
\\
a^{[4]} &= \frac{\beta}{\alpha^{4}} 
\bigg[ 
12 \Big( 
|\u{}{}|^{2}
+ |\u{}{}(-x,-t)|^{2} 
\Big)^{2}
+ 6 \Big( 
|\u{x}{}|^{2}
+ |\u{x}{}(-x,-t)|^{2} 
\Big)
\\&
-2 \Big(
|\u{}{}|^{2}
+ |\u{}{}(-x,-t)|^{2} 
\Big)_{xx}
\bigg], 
\\
a^{[5]} &= 0,
\\
a^{[6]} &= \frac{\beta}{\alpha^{6}} 
\bigg[ 
80 (|\u{}{}|^{2}
+ |\u{}{}(-x,-t)|^{2})^{3}
+80 \Big(|\u{}{}|^{2}
+ |\u{}{}(-x,-t)|^{2} \Big)
\Big( (|\u{x}{}|^{2}
+ |\u{x}{}(-x,-t)|^{2} \Big)
\\&
-40 \Big(|\u{}{}|^{2}
+ |\u{}{}(-x,-t)|^{2} \Big)
\Big( |\u{}{}|^{2}
+ |\u{}{}(-x,-t)|^{2} \Big)_{xx}
\\&
+10 \Big( |\u{xx}{}|^{2}
+ |\u{xx}{}(-x,-t)|^{2} \Big)
+2 \Big( |\u{}{}|^{2}
+ |\u{}{}(-x,-t)|^{2} \Big)_{xxxx}
\\&
-10 \Big( (|\u{}{}|^{2}
+ |\u{}{}(-x,-t)|^{2})_{x} \Big)^{2}
-10 \Big( |\u{x}{}|^{2}
+ |\u{x}{}(-x,-t)|^{2} \Big)_{xx}
\bigg],
\end{cases}
\end{aligned}&
\end{flalign}
\vspace{-2cm}
\begin{flalign}
&\begin{aligned}
\begin{cases}
\b{1}{[0]} &=0 ,
\\
\b{1}{[1]}&=\frac{\beta}{\alpha} \u{}{} , 
\\
\b{1}{[2]} &=- \i \frac{\beta}{\alpha^2} \u{x}{}, 
\\
\b{1}{[3]} &=-\frac{\beta}{\alpha^3} \Big[ 
\u{xx}{} 
-4 \Big( 
|\u{}{}|^{2}+|\u{}{}(-x,-t)|^{2} 
\Big)
\u{}{} \Big], 
\\[3mm]
\b{1}{[4]} &= \i \frac{\beta}{\alpha^{4}}  
\Big[ \u{xxx}{}
-6 \Big(
|\u{}{}|^{2}+|\u{}{}(-x,-t)|^{2} 
\Big)
\u{x}{}
-3 \Big(
|\u{}{}|^{2}+|\u{}{}(-x,-t)|^{2} 
\Big)_{x}
\u{}{}
\Big],
\\
\b{1}{[5]} &= \frac{\beta}{\alpha^{5}}
\Big[ \u{xxxx}{}
-8 \Big( 
(|\u{}{}|^{2}+|\u{}{}(-x,-t)|^{2}) \u{x}{}
\Big)_{x}
\\&
+\Big( 
24 (|\u{}{}|^{2}+|\u{}{}(-x,-t)|^{2})^{2}
+8 (|\u{x}{}|^{2}+|\u{x}{}(-x,-t)|^{2})
-6 (|\u{}{}|^{2}+|\u{}{}(-x,-t)|^{2})_{xx}
\Big) \u{}{}
\\
\b{1}{[6]} &= -\i \frac{\beta}{\alpha^{6}}
\Bigg[
\u{xxxxx}{}
-10 (|u|^{2}+|u(-x,-t)|^{2}) \u{xxx}{}
-15(|u|^{2}+|u(-x,-t)|^{2})_{x} \u{xx}{}
\\ 
&
+ \big[ 
40 (|u|^{2}+|u(-x,-t)|^{2})^{2}
+10 (|\u{x}{}|^{2}+|\u{x}{}(-x,-t)|^{2})
-15 (|u|^{2}+|u(-x,-t)|^{2})_{xx}
\big] \u{x}{}
\\& \label{b1equation}
+ \Big[
20 \Big( (|\u{}{}|^{2}+|\u{}{}(-x,-t)|^{2})^{2} \Big)_{x}
+5 (|\u{x}{}|^{2}+|\u{x}{}(-x,-t)|^{2})_{x}
-5 (|u|^{2}+|\u{}{}(-x,-t)|^{2})_{xxx}
\Big] \u{}{}
\Bigg],
\end{cases}
\end{aligned}&
\end{flalign}
\vspace{-0.5cm}
\begin{flalign}\label{d11-SASA}
&\begin{aligned}
\begin{cases}
\d{11}{[0]} &= \beta_{2},
\\ 
\d{11}{[1]}  &=0, 
\\
\d{11}{[2]}  &= -\frac{\beta}{\alpha^{2}} |\u{}{}|^{2}, 
\\
\d{11}{[3]}  &= 2 \frac{\beta}{\alpha^{3}}
\imag ( \u{}{} \ubar{x}{} ), 
\\
\d{11}{[4]}  &= -\frac{\beta}{\alpha^{4}}
\Big[
6 \big( |\u{}{}|^{2}+|\u{}{}(-x,-t)|^{2} \big) |\u{}{}|^{2}
+3|\u{x}{}|^{2} 
- ( |\u{}{}|^{2} )_{xx}
\Big]
,
\\
\d{11}{[5]}  &= \frac{\beta}{\alpha^{5}}
\Big[ 
16 \big( |\u{}{}|^{2}+|\u{}{}(-x,-t)|^{2} \big)
\imag (\u{}{} \ubar{x}{} )
+2 \imag ( \u{xxx}{} \ubar{}{} - \u{x}{} \ubar{xx}{} )
\Big]
,
\\
\d{11}{[6]}  &= \frac{\beta}{\alpha^{6}}
\Big[ 
\Big( 
10 (|\u{}{}|^{2}+|\u{}{}(-x,-t)|^{2})_{xx}
-10 (|\u{x}{}|^{2}+|\u{x}{}(-x,-t)|^{2})
-40 (|\u{}{}|^{2}+|\u{}{}(-x,-t)|^{2})^{2}
\Big) |\u{}{}|^{2}
\\&
+5 (|\u{}{}|^{2}+|\u{}{}(-x,-t)|^{2})_{x}
(|\u{}{}|^{2})_{x}
+10 \big( |\u{}{}|^{2}+|\u{}{}(-x,-t)|^{2} \big)
\big( (|\u{}{}|^{2})_{xx} -3 |\u{x}{}|^{2} \big)
\\&
-(|\u{}{}|^{2})_{xxxx} 
+5 (|\u{x}{}|^{2})_{xx}
-5 |\u{xx}{}|^{2}
\Big],
\end{cases}
\end{aligned}&
\end{flalign}
\vspace{-0.5cm}
\begin{flalign}\label{d21-SASA}
&\begin{aligned}
\begin{cases}
\d{21}{[0]} &= 0,
\\ 
\d{21}{[1]}  &=0, 
\\
\d{21}{[2]}  &= -\frac{\beta}{\alpha^{2}}
\u{}{}(x,t) \ubar{}{}(-x,-t) , 
\\
\d{21}{[3]}  &= \i \frac{\beta}{\alpha^{3}}
\Big( 
 \ubar{}{}(-x,-t) \u{x}{}(x,t) 
+ \ubar{x}{}(-x,-t) \u{}{}(x,t)
\Big), 
\\
\d{21}{[4]}  &= -\frac{\beta}{\alpha^{4}}
\Big[
6 \big( |\u{}{}|^{2}+|\u{}{}(-x,-t)|^{2} \big) 
\u{}{} \ubar{}{}(-x,-t)
-3 \u{x}{} \ubar{x}{}(-x,-t)
- (\u{}{} \ubar{}{}(-x,-t))_{xx}
\Big]
,
\\
\d{21}{[5]}  &= \i \frac{\beta}{\alpha^{5}}
\Big[
8 \big( |\u{}{}|^{2}+|\u{}{}(-x,-t)|^{2} \big)
(\u{}{} \ubar{x}{}(-x,-t) + \u{x}{} \ubar{}{}(-x,-t))
\\&
-(\u{}{} \ubar{x}{}(-x,-t) + \u{x}{} \ubar{}{}(-x,-t))_{xx}
-2 (\u{x}{} \ubar{xx}{}(-x,-t) + \u{xx}{} \ubar{x}{}(-x,-t))
\Big],
\\
\d{21}{[6]}  &= \frac{\beta}{\alpha^{6}}
\Big[ 
\Big( 
10 (|\u{}{}|^{2}+|\u{}{}(-x,-t)|^{2})_{xx}
-10 (|\u{x}{}|^{2}+|\u{x}{}(-x,-t)|^{2})
\\& 
-40 (|\u{}{}|^{2}+|\u{}{}(-x,-t)|^{2})^{2}
\Big) \u{}{} \ubar{}{}(-x,-t)
+5 (|\u{}{}|^{2}+|\u{}{}(-x,-t)|^{2})_{x}
(\u{}{} \ubar{}{}(-x,-t))_{x}
\\&
+10 \big( |\u{}{}|^{2}+|\u{}{}(-x,-t)|^{2} \big)
\big( (\u{}{} \ubar{}{}(-x,-t))_{xx} 
+3 \u{x}{} \ubar{x}{}(-x,-t) \big)
\\&
-(\u{}{} \ubar{}{}(-x,-t))_{xxxx} 
-5 (\u{x}{} \ubar{x}{}(-x,-t))_{xx}
-5 \u{xx}{} \ubar{xx}{}(-x,-t)
\Big],
\end{cases}
\end{aligned}&
\end{flalign}
\vspace{-0.5cm}
\begin{flalign}\label{d31-SASA}
&\begin{aligned}
\begin{cases}
\d{31}{[0]} &= 0,
\\ 
\d{31}{[1]}  &=0, 
\\
\d{31}{[2]}  &= -\frac{\beta}{\alpha^{2}}
\u{}{2}, 
\\
\d{31}{[3]}  &= 0, 
\\
\d{31}{[4]}  &= -\frac{\beta}{\alpha^{4}}
\Big[
6 \big( |\u{}{}|^{2}+|\u{}{}(-x,-t)|^{2} \big) \u{}{2}
+3 \u{x}{2}
- (\u{}{2})_{xx}
\Big]
,
\\
\d{31}{[5]}  &= 0,
\\
\d{31}{[6]}  &= \frac{\beta}{\alpha^{6}}
\Big[ 
\Big( 
10 (|\u{}{}|^{2}+|\u{}{}(-x,-t)|^{2})_{xx}
-10 (|\u{x}{}|^{2}+|\u{x}{}(-x,-t)|^{2})
-40 (|\u{}{}|^{2}+|\u{}{}(-x,-t)|^{2})^{2}
\Big) \u{}{2}
\\&
+5 (|\u{}{}|^{2}+|\u{}{}(-x,-t)|^{2})_{x}
(\u{}{2})_{x}
+10 \big( |\u{}{}|^{2}+|\u{}{}(-x,-t)|^{2} \big)
\big( (\u{}{2})_{xx} -3 \u{x}{2} \big)
\\&
-(\u{}{2})_{xxxx} 
+5 (\u{x}{2})_{xx}
-5 \u{xx}{2}
\Big],
\end{cases}
\end{aligned}&
\end{flalign}
\vspace{-0.5cm}
\begin{flalign}\label{d41-SASA}
&\begin{aligned}
\begin{cases}
\d{41}{[0]} &= 0,
\\ 
\d{41}{[1]}  &=0, 
\\
\d{41}{[2]}  &= -\frac{\beta}{\alpha^{2}}
\u{}{} \u{}{}(-x,-t), 
\\
\d{41}{[3]}  &= \i \frac{\beta}{\alpha^{3}}
\Big( 
\u{x}{} \u{}{}(-x,-t) 
+ \u{}{} \u{x}{}(-x,-t)
\Big), 
\\
\d{41}{[4]}  &= -\frac{\beta}{\alpha^{4}}
\Big[
6 \big( |\u{}{}|^{2}+|\u{}{}(-x,-t)|^{2} \big) 
\u{}{} \u{}{}(-x,-t)
+3 \u{x}{} \u{x}{}(-x,-t)
+ (\u{}{} \u{}{}(-x,-t))_{xx}
\Big]
,
\\
\d{41}{[5]}  &= \i \frac{\beta}{\alpha^{5}}
\Big[
8 \big( |\u{}{}|^{2}+|\u{}{}(-x,-t)|^{2} \big)
(\u{}{} \u{x}{}(-x,-t) + \u{x}{} \u{}{}(-x,-t))
\\&
-(\u{}{} \u{x}{}(-x,-t) + \u{x}{} \u{}{}(-x,-t))_{xx}
-2 (\u{x}{} \u{xx}{}(-x,-t) + \u{xx}{} \u{x}{}(-x,-t))
\Big],
\\
\d{41}{[6]}  &= \frac{\beta}{\alpha^{6}}
\Big[ 
\Big( 
10 (|\u{}{}|^{2}+|\u{}{}(-x,-t)|^{2})_{xx}
-10 (|\u{x}{}|^{2}+|\u{x}{}(-x,-t)|^{2})
\\& 
-40 (|\u{}{}|^{2}+|\u{}{}(-x,-t)|^{2})^{2}
\Big) \u{}{} \u{}{}(-x,-t)
+5 (|\u{}{}|^{2}+|\u{}{}(-x,-t)|^{2})_{x}
(\u{}{} \u{}{}(-x,-t))_{x}
\\&
+10 \big( |\u{}{}|^{2}+|\u{}{}(-x,-t)|^{2} \big)
\big( (\u{}{} \u{}{}(-x,-t))_{xx} 
+3 \u{x}{} \u{x}{}(-x,-t) \big)
\\&
-(\u{}{} \u{}{}(-x,-t))_{xxxx} 
-5 (\u{x}{} \u{x}{}(-x,-t))_{xx}
-5 \u{xx}{} \u{xx}{}(-x,-t)
\Big],
\end{cases}
\end{aligned}&
\end{flalign}
where $\beta=\beta_{1}-\beta_{2}$.
\begin{remark}
Under the symmetry relations (\ref{Pequ-SASA}), one can show that
$W$ satisfies the equations:
\begin{align}\label{Wequ-SASA}
\begin{cases}
W^{\dagger}(x,t,-\lambar{}{})=
-W(x,t,\lam{}{}),
\\
W^{T}(x,t,-\lam{}{})=C_{1}W(x,t,\lam{}{})C_{1}^{-1},
\\
W^{\dagger}(-x,-t,-\lambar{}{})=
C_{2}W(x,t,\lam{}{})C_{2}^{-1},
\\
\Wbar{}{}(-x,-t,\lambar{}{})=C_{3}W(x,t,\lam{}{})C_{3}^{-1},
\end{cases}
\end{align}
for a solution $W$ to the stationary zero curvature equation.
Using the Laurent expansion (\ref{WLaurentexpansion-SASA}) of $W$, we get the relations:
\begin{align}\label{W-C0-relation-set}
\a{}{[m]}(x,t)&=(-1)^{m+1}\abar{}{[m]}(x,t),
\\
\b{1}{[m]}(x,t)&=(-1)^{m+1} \cbar{1}{[m]}(x,t),
\quad
\b{2}{[m]}(x,t)=(-1)^{m+1} \cbar{2}{[m]}(x,t),
\\
\b{3}{[m]}(x,t)&=(-1)^{m+1} \cbar{3}{[m]}(x,t),
\quad
\b{4}{[m]}(x,t)=(-1)^{m+1} \cbar{4}{[m]}(x,t),
\\
\dbar{11}{[m]}(x,t)&=(-1)^{m+1} \d{11}{[m]}(x,t),
\quad
\dbar{21}{[m]}(x,t)=(-1)^{m+1} \d{12}{[m]}(x,t),
\\
\dbar{31}{[m]}(x,t)&=(-1)^{m+1} \d{13}{[m]}(x,t),
\quad
\dbar{41}{[m]}(x,t)=(-1)^{m+1} \d{14}{[m]}(x,t),
\\
\dbar{22}{[m]}(x,t)&=(-1)^{m+1} \d{22}{[m]}(x,t),
\quad
\dbar{32}{[m]}(x,t)=(-1)^{m+1} \d{23}{[m]}(x,t),
\\
\dbar{42}{[m]}(x,t)&=(-1)^{m+1} \d{24}{[m]}(x,t),
\quad
\dbar{33}{[m]}(x,t)=(-1)^{m+1} \d{33}{[m]}(x,t),
\\
\dbar{43}{[m]}(x,t)&=(-1)^{m+1} \d{34}{[m]}(x,t),
\quad
\dbar{44}{[m]}(x,t)=(-1)^{m+1} \d{44}{[m]}(x,t),
\end{align}
\begin{align}\label{W-C2-relation-set}
\a{}{[m]}(x,t)&=(-1)^{m}\a{}{[m]}(x,t),
\\
\b{1}{[m]}(x,t)&=(-1)^{m} \c{3}{[m]}(x,t),
\quad
\b{2}{[m]}(x,t)=(-1)^{m} \c{4}{[m]}(x,t),
\\
\b{3}{[m]}(x,t)&=(-1)^{m} \c{1}{[m]}(x,t),
\quad
\b{4}{[m]}(x,t)=(-1)^{m} \c{2}{[m]}(x,t),
\\
\d{11}{[m]}(x,t)&=(-1)^{m} \d{33}{[m]}(x,t),
\quad
\d{21}{[m]}(x,t)=(-1)^{m} \d{34}{[m]}(x,t),
\\
\d{31}{[m]}(x,t)&=(-1)^{m} \d{31}{[m]}(x,t),
\quad
\d{41}{[m]}(x,t)=(-1)^{m} \d{32}{[m]}(x,t),
\\
\d{12}{[m]}(x,t)&=(-1)^{m} \d{43}{[m]}(x,t),
\quad
\d{22}{[m]}(x,t)=(-1)^{m} \d{44}{[m]}(x,t),
\\ 
\d{42}{[m]}(x,t)&=(-1)^{m} \d{42}{[m]}(x,t),
\quad
\d{13}{[m]}(x,t)=(-1)^{m} \d{13}{[m]}(x,t),
\\ 
\d{23}{[m]}(x,t)&=(-1)^{m} \d{14}{[m]}(x,t),
\quad
\d{24}{[m]}(x,t)=(-1)^{m} \d{24}{[m]}(x,t).
\end{align}
\vspace{-6mm}
and
\vspace{-6mm}
\begin{align}\label{W-C1-relation-set}
\abar{}{[m]}(-x,-t)&=(-1)^{m}\a{}{[m]}(x,t),
\\
\b{1}{[m]}(-x,-t)&=(-1)^{m} \cbar{2}{[m]}(x,t),
\quad
\b{2}{[m]}(-x,-t)=(-1)^{m} \cbar{1}{[m]}(x,t),
\\
\b{3}{[m]}(-x,-t)&=(-1)^{m} \cbar{4}{[m]}(x,t),
\quad
\b{4}{[m]}(-x,-t)=(-1)^{m} \cbar{3}{[m]}(x,t),
\\
\dbar{11}{[m]}(-x,-t)&=(-1)^{m} \d{22}{[m]}(x,t),
\quad
\dbar{21}{[m]}(-x,-t)=(-1)^{m} \d{21}{[m]}(x,t),
\\
\dbar{31}{[m]}(-x,-t)&=(-1)^{m} \d{24}{[m]}(x,t),
\quad
\dbar{41}{[m]}(-x,-t)=(-1)^{m} \d{23}{[m]}(x,t),
\\
\dbar{12}{[m]}(-x,-t)&=(-1)^{m} \d{12}{[m]}(x,t),
\quad
\dbar{32}{[m]}(-x,-t)=(-1)^{m} \d{14}{[m]}(x,t),
\\
\dbar{42}{[m]}(-x,-t)&=(-1)^{m} \d{13}{[m]}(x,t),
\quad
\dbar{33}{[m]}(-x,-t)=(-1)^{m} \d{44}{[m]}(x,t),
\\
\dbar{43}{[m]}(-x,-t)&=(-1)^{m} \d{43}{[m]}(x,t),
\quad
\dbar{34}{[m]}(-x,-t)=(-1)^{m} \d{34}{[m]}(x,t),
\end{align}
\vspace{-6mm}
and finally
\vspace{-6mm}
\begin{align}
\a{}{[m]}(x,t)&=\abar{}{[m]}(-x,-t),
\\
\b{1}{[m]}(x,t)&= \bbar{4}{[m]}(-x,-t),
\quad
\b{2}{[m]}(x,t)= \bbar{3}{[m]}(-x,-t),
\\
\c{1}{[m]}(x,t)&= \cbar{4}{[m]}(-x,-t),
\quad
\c{2}{[m]}(x,t)= \cbar{3}{[m]}(-x,-t),
\\
\d{11}{[m]}(x,t)&= \dbar{44}{[m]}(-x,-t),
\quad
\d{12}{[m]}(x,t)= \dbar{43}{[m]}(-x,-t),
\\
\d{13}{[m]}(x,t)&= \dbar{42}{[m]}(-x,-t),
\quad
\d{14}{[m]}(x,t)= \dbar{41}{[m]}(-x,-t),
\\
\d{21}{[m]}(x,t)&= \dbar{34}{[m]}(-x,-t),
\quad
\d{22}{[m]}(x,t)=  \dbar{33}{[m]}(-x,-t),
\\ 
\d{23}{[m]}(x,t)&= \dbar{32}{[m]}(-x,-t),
\quad
\d{24}{[m]}(x,t)= \dbar{31}{[m]}(-x,-t).
\label{W-C3-relation-set}
\end{align}
\end{remark}
\noindent
One can easily relate all the members of the set 
$\mathcal{S}_{1}=\{ \b{1}{[m]},\b{2}{[m]},\b{3}{[m]},\b{4}{[m]},\c{1}{[m]},\c{2}{[m]},\c{3}{[m]},\c{4}{[m]}\}$
directly from the above four sets of relations.
Similarly, the members within each of the following sets 
$\mathcal{S}_{2}=\{ \d{11}{[m]},\d{22}{[m]}, \d{33}{[m]}, \d{44}{[m]} \}$, 
$\mathcal{S}_{3}=\{ \d{21}{[m]}, \d{34}{[m]}, \d{12}{[m]} ,\d{43}{[m]}\}$, 
$\mathcal{S}_{4}=\{ \d{31}{[m]}, \d{24}{[m]}, \d{13}{[m]}, \d{42}{[m]}\}$
and $\mathcal{S}_{5}=\{ \d{41}{[m]}, \d{23}{[m]}, \d{32}{[m]}, \d{14}{[m]} \}$ are related to each other. As a consequence, any member in the set $\mathcal{S}_{1}$ can be expressed in terms of the independents
$\{ \a{}{[m]}, \d{11}{[m]}, \d{21}{[m]}, \d{31}{[m]}, \d{41}{[m]} \}$.
\newline
We introduce the Lax matrix
\begin{align*}
V^{[m]}(\u{}{},\lambda)&=(\lambda^{m} W)_{+}
= \sum\limits_{i=0}^{m}W_{i}\lambda^{m-i}
= \sum\limits_{i=0}^{m}
\begin{pmatrix} 
a^{[i]} \lam{}{m-i}    & 
\b{1}{[i]} \lam{}{m-i} & 
\b{2}{[i]} \lam{}{m-i} & 
\b{3}{[i]} \lam{}{m-i} & 
\b{4}{[i]} \lam{}{m-i}
\\
\c{1}{[i]}  \lam{}{m-i} & 
\d{11}{[i]} \lam{}{m-i} & 
\d{12}{[i]} \lam{}{m-i} & 
\d{13}{[i]} \lam{}{m-i} & 
\d{14}{[i]} \lam{}{m-i} 
\\
\c{2}{[i]} \lam{}{m-i}  & 
\d{21}{[i]} \lam{}{m-i} & 
\d{22}{[i]} \lam{}{m-i} & 
\d{23}{[i]} \lam{}{m-i} & 
\d{24}{[i]} \lam{}{m-i}
\\
\c{3}{[i]}  \lam{}{m-i}  & 
\d{31}{[i]} \lam{}{m-i}  & 
\d{32}{[i]} \lam{}{m-i}  & 
\d{33}{[i]} \lam{}{m-i}  & 
\d{34}{[i]} \lam{}{m-i}
\\
\c{4}{[i]}  \lam{}{m-i}  & 
\d{41}{[i]} \lam{}{m-i}  & 
\d{42}{[i]} \lam{}{m-i}  & 
\d{43}{[i]} \lam{}{m-i}  & 
\d{44}{[i]} \lam{}{m-i}
\end{pmatrix},
\end{align*}
where the modification terms are taken to be zero . Therefore, we get the spatial and temporal equations of the spectral problems \cite{Ma2018}, with the associated Lax pair $\{U,V^{[m]}\}$: 
\begin{align}\label{AKNSsystem1-SASA}
\psi_{x} &= \i U \psi,
\\ \label{AKNSsystem2-SASA}
\psi_{t} &= \i {V^{[m]}} \psi.
\end{align}
The compatibility
conditions of equations (\ref{AKNSsystem1-SASA})-(\ref{AKNSsystem2-SASA})
give rise to the zero curvature equations
\begin{equation}
ZCE:= U_{t_{m}} - V_{x}^{[m]} + \i [U,V^{[m]}] = 0.
\end{equation}
The comptabiltiy of the second component
in the first row and the fourth component of the first column of $ZCE$, namely $(ZCE)_{12}$ and $(ZCE)_{41}$ lead to the  scalar Sasa-Satsuma integrable hierarchy:
\begin{align}
\u{t_{m}}{}&=
\begin{cases}
\i \alpha 
\b{1}{[m+1]},
\quad m=\text{odd},
\\
0, \quad m=\text{even},
\end{cases}
\quad m \geq 0.
\end{align}
Obtaining a  hierarchy that only generates 
mKdV-type equations, but not NLS-type equations. More specifically, the hierarchy here gives Sasa-Satsuma-type equations due to the initial choice of the matrix $U(u,\lam{}{})$. Thus,
\begin{equation}\label{AKNShierarchy-SASA}
\u{t_{m}}{}(x,t)=\i \alpha 
\b{1}{[m+1]},
\quad m=\text{odd}.
\end{equation}
For example, the case of $m=3$ leads to the nonlocal reverse-spacetime Sasa-Satsuma equation \cite{MaMathematics2022}:
\begin{flalign}\label{Sasa-Satsuma-coupled1-SASA}
\u{t}{}(x,t) + \frac{\beta}{\alpha^{3}}  
\Big[ \u{xxx}{}
-6 \Big(
|\u{}{}|^{2}+|\u{}{}(-x,-t)|^{2} 
\Big)
\u{x}{}
-3 \Big(
|\u{}{}|^{2}+|\u{}{}(-x,-t)|^{2} 
\Big)_{x}
\u{}{}
\Big]=0.
\end{flalign}
This soliton hierarchy possesses a bi-Hamiltonian structure
\begin{equation}\label{AKNShierarchy2-SASA}
\u{t_{m}}{}=\i \alpha \b{1}{[2m]}
=J_{1} \frac{\delta \mathcal{H}_{2m-1}}{\delta \u{}{}}
=J_{2} \frac{\delta \mathcal{H}_{2m-3}}{\delta \u{}{}},
\quad 
m \in \{2,3,\ldots \},
\end{equation}
where $\mathcal{H}_{2m-3}$ are the Hamiltonian functionals and $J_{1}$ and $J_{2}$ are a Hamiltonian pair .
\newline
We derive from the recursive relations 
(\ref{amrecursiverelation})-(\ref{d43d44recursiverelation}) and the relations 
(\ref{W-C0-relation-set})-(\ref{W-C3-relation-set}),
the following recursive formula between 
$\b{1}{[m+1]}$ and
$\b{1}{[m]}$:
\begin{equation}\label{cmpsirelation-SASA}
\b{1}{[m+1]}
=\Psi 
\b{1}{[m]},
\end{equation}
where the recursion operator $\Psi$ reads:
\begin{align}\nonumber
\Psi &= \frac{\i}{\alpha}
\Big[ -\partial
\\& \nonumber
+\Big( 
(2+(-1)^{m})\u{}{} \dinv{} \ubar{}{} 
+ \u{}{}(-x) \dinv{} \ubar{}{}(-x)
+(1+(-1)^{m}) \ubar{}{} \dinv{} \u{}{} 
+ \ubar{}{}(-x) \dinv{} \u{}{}(-x)
\Big) \Gammaop{+}{}
\\& \nonumber
+\Big( 
(2(-1)^{m+1}-1) \u{}{} \dinv{} \u{}{} 
\Big) \Gammabarop{+}{}
+\Big( 
((-1)^{m+1}-1) \u{}{} \dinv{} \ubar{}{}(-x)
+ (-1)^{m+1} \ubar{}{}(-x) \dinv{} \u{}{} 
\Big) \Gammaop{-}{}
\\&
+\Big( 
((-1)^{m}+1) \u{}{} \dinv{} \u{}{}(-x)
+ (-1)^{m} \u{}{}(-x) \dinv{} \u{}{} 
\Big) \Gammabarop{-}{}
\Big],
\end{align}
and where the operators $\Gammaop{\pm}{},\Gammabarop{\pm}{}$
are defined by:
\begin{align} 
\Gammaop{-}{} f(x,t) &= f(-x,-t),
\\
\Gammabarop{+}{} f(x,t) &= \overset{*}{f}(x,t),
\\
\Gammabarop{-}{} f(x,t) &= \overset{*}{f}(-x,-t),
\end{align}
with $\Gammaop{+}{}$ being the identity operator, i.e.,
$\Gammaop{+}{} f(x,t)=\Identityop f(x,t)=f(x,t)$.
\subsection{Nonlocal reverse-spacetime Sasa-Satsuma equation}
\noindent
To derive the one-component nonlocal 
Sasa-Satsuma equation, we take the Lax matrix
\begin{equation}
V^{[5]}=V^{[5]}(\u{}{},\lam{}{})=(\lam{}{5} W)_{+}.
\end{equation}
The spatial and temporal equations of the spectral problems
(\ref{AKNSsystem1-SASA}) and (\ref{AKNSsystem2-SASA}), with the associated 
Lax pair $\{U,V^{[5]}\}$ read,
\begin{align}\label{AKNSsystem-SASA}
\psi_{x} &= \i U \psi,
\\
\psi_{t} &= \i V^{[5]} \psi,
\end{align}
with the zero curvature equation
\begin{equation}
U_{t} - V_{x}^{[5]} + \i [U,V^{[5]}] = 0 ,
\end{equation}
that gives the scalar Sasa-Satsuma equation
\begin{align}\label{solitonequation-SASA}
\u{t}{}(x,t)
=\i \alpha 
\b{1}{[6]}.
\end{align}
Explicitly,
\begin{flalign}\label{Sasa-Satsuma-equation-u-SASA-2}
\nonumber
\u{t}{}&=\u{xxxxx}{}
-10 (|u|^{2}+|u(-x,-t)|^{2}) \u{xxx}{}
-15(|u|^{2}+|u(-x,-t)|^{2})_{x} \u{xx}{}
\\ \nonumber
&
+ \big[ -15 (|u|^{2}+|u(-x,-t)|^{2})_{xx}
+10 (|\u{x}{}|^{2}+|\u{x}{}(-x,-t)|^{2})
+ 40 (|u|^{2}+|u(-x,-t)|^{2})^{2} \big] \u{x}{}
\\& 
+ \big[
-5 (|u|^{2}+|\u{}{}(-x,-t)|^{2})_{xxx}
+5 (|\u{x}{}|^{2}+|\u{x}{}(-x,-t)|^{2})_{x}
+20 \big( (|u|^{2}+|\u{}{}(-x,-t)|^{2})^{2} \big)_{x}
\big] \u{}{},
\end{flalign}
and
\begin{equation}
V^{[5]}=
\begin{pmatrix}
V_{11} & V_{12} & V_{13} & V_{14} & V_{15}
\\
V_{21} & V_{22} & V_{23} & V_{24} & V_{25}
\\
V_{31} & V_{32} & V_{33} & V_{34} & V_{35}
\\
V_{41} & V_{42} & V_{43} & V_{44} & V_{45}
\\
V_{51} & V_{52} & V_{53} & V_{54} & V_{55}
\end{pmatrix},
\end{equation}
where the components are explicitly 
\begin{align*}
V_{11} &= 
\sum\limits_{i=0}^{2}
\lam{}{5-2i} \a{}{[2i]}, 
&
V_{12} &=
\sum\limits_{i=0}^{5}
\lam{}{[5-i]} \b{1}{[i]},
&
V_{13} &=
- \sum\limits_{i=0}^{5}
\lam{}{5-i} \Gammaop{-}{} \b{1}{[i]},
\\
V_{21} &=
\sum\limits_{i=0}^{5}
(-1)^{i+1} \lam{}{5-i} \Gammabarop{+}{} 
\b{1}{[i]},
&
V_{22} &=
\sum\limits_{i=0}^{5}
\lam{}{5-i} \d{11}{[i]},
&
V_{23} &=
\sum\limits_{i=0}^{5}
(-1)^{i+1} \lam{}{5-i} 
\Gammabarop{+}{} \d{21}{[i]},
\\
V_{31} &=
\sum\limits_{i=0}^{5}
(-1)^{i} \lam{}{5-i} \Gammabarop{-}{} 
\b{1}{[i]},
&
V_{32} &=
\sum\limits_{i=0}^{5}
\lam{}{5-i} \d{21}{[i]},
&
V_{33} &=
\sum\limits_{i=0}^{5}
(-1)^{i} \lam{}{5-i} 
\Gammabarop{-}{} \d{11}{[i]},
\\
V_{41} &=
\sum\limits_{i=0}^{5}
(-1)^{i} \lam{}{5-i} 
\b{1}{[i]},
&
V_{42} &=
\sum\limits_{i=0}^{2}
\lam{}{5-2i} \d{31}{[2i]},
&
V_{43} &=
- \sum\limits_{i=0}^{5}
\lam{}{5-i} \Gammaop{-}{} \d{41}{[i]},
\\
V_{51} &=
\sum\limits_{i=0}^{5}
(-1)^{i+1} \lam{}{5-i} \Gammaop{-}{} 
\b{1}{[i]},
&
V_{52} &=
\sum\limits_{i=0}^{5}
\lam{}{5-i} \d{41}{[i]},
&
V_{53} &=
- \sum\limits_{i=0}^{2}
\lam{}{5-2i} \Gammaop{-}{} \d{31}{[2i]},
\end{align*}
\vspace{-3mm}
\begin{align*}
V_{14} &=
- \sum\limits_{i=0}^{5}
\lam{}{5-i} \Gammabarop{+}{} \b{1}{[i]},
&
V_{15} &=
\sum\limits_{i=0}^{5}
\lam{}{5-i} \Gammabarop{-}{} \b{1}{[i]},
\\
V_{24} &=
\sum\limits_{i=0}^{2}
(-1)^{2i+1} \lam{}{5-2i} \Gammabarop{+}{} \d{31}{[2i]},
&
V_{25} &=
\sum\limits_{i=0}^{5}
(-1)^{i+1} \lam{}{5-i} \Gammabarop{+}{} \d{41}{[i]},
\\
V_{34} &=
- \sum\limits_{i=0}^{5}
\lam{}{5-i} \Gammabarop{+}{} \d{41}{[i]},
&
V_{35} &=
\sum\limits_{i=0}^{2}
(-1)^{2i} \lam{}{5-2i} \Gammabarop{-}{} \d{31}{[2i]},
\\
V_{44} &=
\sum\limits_{i=0}^{5}
(-1)^{i} \lam{}{5-i} \d{11}{[i]},
&
V_{45} &=
\sum\limits_{i=0}^{5}
\lam{}{5-i} \Gammabarop{-}{} \d{21}{[i]},
\\
V_{54} &=
- \sum\limits_{i=0}^{5}
\lam{}{5-i} \Gammabarop{+}{} \d{21}{[i]},
&
V_{55} &=
\sum\limits_{i=0}^{5}
\lam{}{5-i} \Gammabarop{-}{} \d{11}{[i]}.
\end{align*}
The matrix $V^{[5]}$ exhibits the properties of symmetry:
\begin{align}\label{Vreduction-SASA}
\begin{cases}
V^{[5] \dagger}(x,t,-\lambar{}{})=V^{[5]}(x,t,\lambda),
\\
V^{[5] T}(x,t,-\lam{}{})=-C_{1}V^{[5]}(x,t,\lam{}{})C_{1}^{-1},
\\
V^{[5] \dagger}(-x,-t,-\lambar{}{})=-C_{2}V^{[5]}(x,t,\lambda)C_{2}^{-1},
\\
\Vbar{}{[5]}(-x,-t,\lambar{}{})=C_{3}V^{[5]}(x,t,\lambda)C_{3}^{-1}.
\end{cases}
\end{align}
\subsection{Bi-Hamiltonian Structures}
\noindent
We start to find a bi-Hamiltonian structures of the soliton hierarchy (\ref{AKNShierarchy-SASA}). To do so, we are going to use the trace identity 
\begin{equation}\label{traceidentity-SASA}
\frac{\delta}{\delta \u{}{}}
\int tr \bigg[ W \frac{\partial U}{\partial \lambda} \bigg] dx
= \lambda^{- \gamma}
\frac{\partial}{\partial \lambda}
\bigg[ \lambda^{\gamma}
tr \big( W \frac{\partial U}{\partial \u{}{}} \big)
\bigg],
\end{equation}
where
\begin{equation}
\gamma=-\frac{\lambda}{2} \frac{d}{d\lambda}
\ln|tr\big(W^{2}\big)|.    
\end{equation}
Thus, from the matrix $U$, one can easily compute 
$\frac{\partial U}{\partial \u{}{}}$ to obtain,
\begin{align*}
&tr\bigg[ W \frac{\partial U}{\partial \lambda} \bigg]=
\sum\limits_{m=0}^{\infty}
\Big( \alpha_{1} a^{[m]}
+ \alpha_{2} ((-1)^{m}+1) 
(\Gammaop{+}{} + \Gammabarop{-}{}) 
\d{11}{[m]} 
\Big) \lam{}{-m},
\\
&tr \bigg[ W \frac{\partial U}{\partial \u{}{}(x,t)} \bigg]=
\sum\limits_{m=0}^{\infty}
((-1)^{m+1}+1)
\Gammabarop{+}{} \b{1}{[m]}(x,t) \lam{}{-m}.
\end{align*}
By substituting these in the trace identity formula 
(\ref{traceidentity-SASA}), and matching the powers of 
$\lambda^{-m-1}$, we get
\begin{equation}\label{VariHam-SASA}
\frac{\delta}{\delta \u{}{}} \int
\Big( \alpha_{1} a^{[m+1]}
+ \alpha_{2} ((-1)^{m+1}+1) 
(\Gammaop{+}{} + \Gammabarop{-}{}) 
\d{11}{[m+1]} 
\Big) dx
=
(\gamma-m)
((-1)^{m+1}+1)
\Gammabarop{+}{} \b{1}{[m]}
, \, m \geq 1.
\end{equation}
Observing when $m=1$ and $m=3$, we see that $\gamma=2m$.
Hence, the Hamiltonians are given by
\begin{align}\label{Hamiltonian-SASA}
\mathcal{H}_{m}&=
\frac{1}{m}
\int
\Big( \alpha_{1} a^{[m+1]}
+ \alpha_{2} ((-1)^{m+1}+1) 
(\Gammaop{+}{} + \Gammabarop{-}{}) 
\d{11}{[m+1]} 
\Big) dx
\quad m\geq 1
\\&=
\begin{cases}
\frac{1}{m} \int
\Big( \alpha_{1} a^{[m+1]}
+ 2 \alpha_{2}  
(\Gammaop{+}{} + \Gammabarop{-}{}) 
\d{11}{[m+1]} 
\Big) dx,
\quad m=\text{odd},
\\
0, \quad m=\text{even}.
\end{cases}
\\&=
\begin{cases}
\frac{\alpha}{m} \int
a^{[m+1]} dx,
\quad m=\text{odd},
\\
0, \quad m=\text{even},
\end{cases}
\end{align}
since $a^{[m+1]} = -((-1)^{m+1}+1)   
(\Gammaop{+}{} + \Gammabarop{-}{}) 
\d{11}{[m+1]}$ from (\ref{amrecursiverelation}),
(\ref{d11mrecursiverelation})-(\ref{d22mrecursiverelation})
and finally
(\ref{d33mrecursiverelation})-(\ref{d43d44recursiverelation}).
Also, we have
\begin{equation}
\frac{\delta \mathcal{H}_{2m-1}}{\delta \u{}{}} =
2 \Gammabarop{+}{} \b{1}{[2m-1]},
\quad 
m \in \{1,3,\ldots\}
\end{equation}
Now since
\begin{align}
\u{t_{m}}{} &=\i \alpha 
\b{1}{[2m]} 
=J_{1} \frac{\delta \mathcal{H}_{2m-1}}{\delta \u{}{}}
= J_{1} 2 \Gammabarop{+}{} \b{1}{[2m-1]}
\\&
= J_{2} \frac{\delta \mathcal{H}_{2m-3}}{\delta \u{}{}}
=J_{2} 2 \Gammabarop{+}{} \b{1}{[2m-3]},
\quad
\text{for}
\quad
m \in \{2,3,\ldots \}.
\end{align}
where $\Gammabarop{\pm}{}(c \Gammabarop{\pm}{} f)= \cbar{}{} f$, and $c$ is any complex number,
therefore, we deduce the Hamiltonian pair $J_{1}$ and $J_{2}$ as follows:
\begin{equation}
J_{1}=
\i \frac{\alpha}{2} \Psi_{o} \Gammabarop{+}{} 
\end{equation}
and
\begin{equation}
J_{2}=J_{1} \Gammabarop{+}{} \Psi_{e} \Psi_{o}
\Gammabarop{+}{}
\end{equation}
where $\Psi_{o} = \Psi \big|_{\substack{m=odd}}$
and 
$\Psi_{e} = \Psi \big|_{\substack{m=even}}$.
Using the matrix $U$, we can find the first three Hamiltonian functionals:
\begin{align}\label{generalHamiltonians-SASA}
\mathcal{H}_{1}&=2 \frac{\beta}{\alpha}
\int \Big( |\u{}{}(x,t)|^{2}+|\u{}{}(-x,-t)|^{2}
\Big) dx,
\\ \nonumber
\mathcal{H}_{3}&=
\frac{2 \beta}{3 \alpha^{3}}
\int 
\Big[
6 \Big( 
|\u{}{}(x,t)|^{2}+|\u{}{}(-x,-t)|^{2}
\Big)^{2}
+3 
\Big( 
|\u{x}{}(x,t)|^{2}+|\u{x}{}(-x,-t)|^{2}
\Big) 
\\&
- \Big( 
|\u{}{}(x,t)|^{2}+|\u{}{}(-x,-t)|^{2}
\Big)_{xx}
\Big]
dx,
\\ 
\mathcal{H}_{5}&= \nonumber
\frac{\beta}{5 \alpha^{5}} 
\int \bigg[ 
80 (|\u{}{}|^{2}
+ |\u{}{}(-x,-t)|^{2})^{3}
+80 \Big(|\u{}{}|^{2}
+ |\u{}{}(-x,-t)|^{2} \Big)
\Big( (|\u{x}{}|^{2}
+ |\u{x}{}(-x,-t)|^{2} \Big)
\\& \nonumber
-40 \Big(|\u{}{}|^{2}
+ |\u{}{}(-x,-t)|^{2} \Big)
\Big( |\u{}{}|^{2}
+ |\u{}{}(-x,-t)|^{2} \Big)_{xx}
\\& \nonumber
+10 \Big( |\u{xx}{}|^{2}
+ |\u{xx}{}(-x,-t)|^{2} \Big)
+2 \Big( |\u{}{}|^{2}
+ |\u{}{}(-x,-t)|^{2} \Big)_{xxxx}
\\&
-10 \Big( (|\u{}{}|^{2}
+ |\u{}{}(-x,-t)|^{2})_{x} \Big)^{2}
-10 \Big( |\u{x}{}|^{2}
+ |\u{x}{}(-x,-t)|^{2} \Big)_{xx}
\bigg] dx.
\end{align}
\section{Riemann-Hilbert problems}
\noindent
The spatial and temporal spectral problems of the two-component nonlocal Sasa-Satsuma equation can be written as:
\begin{equation}\label{spatialequ-SASA}
\psi_{x}=i U \psi = i (\lambda \Lambda + P) \psi,
\end{equation}
\begin{equation}\label{temporalequ-SASA}
\psi_{t}=i V^{[5]} \psi = i (\lambda^{5} \Omega + Q) \psi,
\end{equation}
where 
$\Lambda=\textit{diag}(\alpha_{1},\alpha_{2} I_{4})$, 
$\Omega=\textit{diag}(\beta_{1},\beta_{2} I_{4})$,
and 
$Q=V^{[5]}-\lam{}{5}\Omega$.
Throughout the presentation of this paper, we assume that 
$\alpha=\alpha_{1}-\alpha_{2}<0$ and \\ 
$\beta=\beta_{1}-\beta_{2}<0$, where $\beta_{1}+4\beta_{2}=0$.
\newline
To find soliton solutions, we begin with an initial condition 
$\u{}{}(x,0)$ and evolute in time to reach $\u{}{}(x,t)$. 
Assume that $\u{}{}$ decays exponentially,  i.e., 
$\u{}{} \rightarrow 0$ as
$x,t \rightarrow \pm \infty$.
Therefore from the spectral problems (\ref{spatialequ-SASA})
and (\ref{temporalequ-SASA}), the asymptotic behaviour of the fundamental matrix $\psi$ 
can be written as follows
\begin{equation}
\psi(x,t) \leadsto e^{i \lambda \Lambda x + i \lam{}{5}\Omega t}. 
\end{equation}
Hence, the solution of the spectral problems can be written in the form:
\begin{equation}\label{psiequ-SASA}
\psi(x,t) = \phi(x,t) e^{i \lambda \Lambda x +i \lam{}{5}\Omega t}.
\end{equation}
The Jost solution of the eigenfunction (\ref{psiequ-SASA}) requires that \cite{Yang2010,DrazinJohnson1989}
\begin{equation}\label{boundary-SASA}
\quad \phi(x,t) \rightarrow I_{5},\quad \text{as}\quad x,t \rightarrow \pm \infty,  
\end{equation}
where $I_{5}$ is the $5 \times 5$ identity matrix. We denote
\begin{equation}
\phi^{\pm} \rightarrow I_{5}, \quad \text{when} \quad x \rightarrow \pm \infty.
\end{equation}
Using equation (\ref{psiequ-SASA}), the Lax pair (\ref{spatialequ-SASA}) and (\ref{temporalequ-SASA}) can be rewritten in terms of $\phi$ so that the spectral problems can be written equivalently as:
\begin{equation}\label{spatialphi-SASA}
\phi_{x} = i \lambda [\Lambda,\phi] +i P \phi,
\end{equation}
\begin{equation}\label{temporalphi-SASA}
\phi_{t} = i \lam{}{5} [\Omega,\phi] +i Q \phi.
\end{equation}
To construct the Riemann-Hilbert problems and their solutions in the reflectionless case, we are going to use the adjoint scattering equations of the spectral problems 
$\psi_{x}=i U \psi$ and $\psi_{t}=i V^{[5]} \psi$. 
Their adjoints are given by
\begin{equation}\label{psiadjointequation-SASA}
\tilde{\psi}_{x} = -i \tilde{\psi} U,
\end{equation}
\begin{equation}\label{psiadjointtemporalequation-SASA}
\tilde{\psi}_{t} = -i \tilde{\psi} V^{[5]},
\end{equation}
and the equivalent spectral adjoint equations read
\begin{equation}\label{adjointspatialphi-SASA}
\tilde{\phi}_{x} = -i \lam{}{} [\tilde{\phi},\Lambda]-i\tilde{\phi} P,
\end{equation}
\begin{equation}\label{adjointtemporalphi-SASA}
\tilde{\phi}_{t} = - i \lam{}{5} [\tilde{\phi},\Omega]-i\tilde{\phi} Q.
\end{equation}
Because $tr(iP)=0$ and $tr(iQ)=0$, using Liouville's formula \cite{Yang2010}, it is easy to see that the $(det(\phi))_{x}=0$, 
that is, $det(\phi)$ is a constant, and utilizing the boundary condition (\ref{boundary-SASA}), we conclude that
\begin{equation}
det(\phi)=1,
\end{equation}
and hence the Jost matrix $\phi$ is invertible.
\newline
Furthermore, as $\phi^{-1}_{x}=-\phi^{-1} \phi_{x} \phi^{-1}$, 
we can derive from (\ref{spatialphi-SASA}),
\begin{equation}\label{adjointspatialphiinv-SASA}
\phi^{-1}_{x} = -i \lambda [\phi^{-1},\Lambda]-i\phi^{-1} P.
\end{equation}
Thus, we can see that both $(\phi^{+})^{-1}$
and $(\phi^{-})^{-1}$
satisfies the spatial adjoint equation (\ref{adjointspatialphi-SASA}). 
We can also show that both satisfies
the temporal adjoint equation (\ref{adjointtemporalphi-SASA}) as well.
\newline
It can be shown that if the eigenfunction $\phi(x,t,\lambda)$ is a solution to the spectral problem (\ref{spatialphi-SASA}), then $\phi^{-1}(x,t,\lambda)$ is a solution to the adjoint spectral problem (\ref{adjointspatialphi-SASA}).
\\[3mm]
This implies that $C_{1} \phi^{-1}(x,t,\lam{}{})$
is also a solution of (\ref{adjointspatialphi-SASA}) with the same eigenvalue, because $\phi_{x}^{-1}=- \phi^{-1} \phi_{x} \phi^{-1}$. 
In a similar way, the nonlocal
$\phi^{T}(x,t,-\lam{}{}) C_{1}$ is also a solution of the spectral adjoint problem (\ref{adjointspatialphi-SASA}).
Since the boundary condition is the same for both solutions as $x \rightarrow \pm \infty$,
this guarantees the uniqueness of the solution, so
\begin{equation}\label{phiTphiminus-SASA}
\phi^{T}(x,t,-\lam{}{}) =
C_{1} \phi^{-1}(x,t,\lam{}{}) C_{1}^{-1}.
\end{equation}
As a result, if $\lambda$
is an eigenvalue of equation (\ref{spatialphi-SASA}) or (\ref{adjointspatialphi-SASA}), then 
$-\lam{}{}$ is also an eigenvalue and the relation
(\ref{phiTphiminus-SASA}) is satisfied.
\newline
In the same way, one can prove that
$\phi^{\dagger}(-x,-t,-\lambar{}{}) C_{2}$ and
$C_{2} \phi^{-1}(x,t,\lambda)$ 
satisfy (\ref{adjointspatialphi-SASA}), while
$\phibar{}{}(-x,-t,\lambar{}{}) C_{3}$
and
$C_{3} \phi(x,t,\lam{}{}) $ 
satisfy (\ref{spatialphi-SASA}).
Thus, using the boundary condition and by uniqueness of the solution, we can also derive
\begin{align}\label{phiTphiminusC2-SASA}
\phi^{\dagger}(-x,-t,-\lambar{}{}) &=
C_{2} \phi^{-1}(x,t,\lambda) C_{2}^{-1},
\\ \label{phiTphiminusC3-SASA}
\phibar{}{}(-x,-t,\lambar{}{}) &=
C_{3} \phi(x,t,\lambda) C_{3}^{-1}.
\end{align}
Now, we are going to work with the spatial spectral problem (\ref{spatialphi-SASA}), assuming that the time is $t=0$.
For notation simplicity, we denote 
$Y^{+}$ and $Y^{-}$ to indicate the boundary conditions are set as $x \rightarrow \infty$
and $x \rightarrow -\infty$, respectively. 
\newline
We know that
\begin{equation}\label{phiboundary-SASA}
\phi^{\pm} \rightarrow I_{5} \quad \text{when} \quad x \rightarrow \pm \infty.
\end{equation}
From (\ref{psiequ-SASA}), we can write
\begin{equation}\label{psiphirelation-SASA}
\psi^{\pm} = \phi^{\pm} e^{i \lambda \Lambda x}.
\end{equation}
Both $\psi^{+}$ and $\psi^{-}$ satisfy 
the spectral spatial differential equation (\ref{spatialequ-SASA}),
i.e. both are two solutions of that equation. 
Thus, they are linearly dependent. So, there exists
a scattering matrix $S(\lambda)$ such that
\begin{equation}\label{psiminusplus-SASA}
\psi^{-} = \psi^{+} S(\lambda),
\end{equation}
and substituting (\ref{psiphirelation-SASA}) into (\ref{psiminusplus-SASA}),
leads to
\begin{equation}\label{psiplusminus-SASA}
\phi^{-} = \phi^{+} e^{i \lambda \Lambda x} S(\lambda) e^{-i \lambda \Lambda x}, \quad \text{for}
\quad \lambda \in \mathbb{R}, 
\end{equation}
where 
\begin{equation}
S(\lambda)=(s_{ij})_{5 \times 5}=
\begin{pmatrix}
s_{11} & s_{12} & s_{13} & s_{14} & s_{15}  \\
s_{21} & s_{22} & s_{23} & s_{24} & s_{25}  \\
s_{31} & s_{32} & s_{33} & s_{34} & s_{35}  \\
s_{41} & s_{42} & s_{43} & s_{44} & s_{45}  \\
s_{41} & s_{42} & s_{43} & s_{44} & s_{55}
\end{pmatrix}.
\end{equation}
Given that $det(\phi^{\pm})=1$, we obtain
\begin{equation}
det(S(\lambda))=1.
\end{equation}
In addition, we can show from
(\ref{phiTphiminus-SASA})-(\ref{phiTphiminusC3-SASA}) and (\ref{psiplusminus-SASA})
that $S(\lambda)$ possesses the involution relations
\begin{align}\label{Sequ-SASA}
S^{T}(-\lam{}{}) &= C_{1} S^{-1}(\lam{}{}) C_{1}^{-1},
\\ \label{SequC2-SASA}
S^{\dagger}(-\lambar{}{}) &= C_{2} S^{-1}(\lambda) C_{2}^{-1},
\\ \label{SequC3-SASA}
\Sbar{}{}(\lambar{}{}) &= C_{3} S(\lambda) C_{3}^{-1}.
\end{align}
We deduce from (\ref{Sequ-SASA})-(\ref{SequC3-SASA}) that
\begin{align}\label{s11hats11relation-SASA}
\hat{s}_{11}(\lam{}{})&=s_{11}(-\lam{}{}),
\\ \label{s11hats11relationC2-SASA}
\hat{s}_{11}({\lambda})&=\sbar{11}{}(-\lambar{}{}),
\\ \label{s11hats11relationC3-SASA}
\sbar{11}{}(\lambar{}{}) &= \s{11}{} (\lam{}{}),
\end{align}
where the inverse of the scattering data matrix is denoted by
$S^{-1}=(\hat{s}_{ij})_{5 \times 5}$ 
for $i,j \in \{1,\ldots,5\}$. 
In order to formulate Riemann-Hilbert problems, we need to analyse the analyticity of the Jost matrix $\phi^{\pm}$. 
Our solutions $\phi^{\pm}$ to this problem can be uniquely written by using the Volterra integral equations in conjunction with the spatial spectral problem (\ref{spatialequ-SASA}):
\begin{align}\label{Volt1-SASA}
\phi^{-}(x,\lambda) &= I_{5} + i \int\limits^{x}_{- \infty} 
e^{i \lambda  (x-y) \Lambda} P(y) \phi^{-}(y,\lambda) 
e^{i \lambda  (y-x) \Lambda} dy,
\\
\phi^{+}(x,\lambda) &= I_{5} - i \int\limits^{+ \infty}_{x} 
e^{i \lambda  (x-y) \Lambda} P(y) \phi^{+}(y,\lambda) 
e^{i \lambda  (y-x) \Lambda} dy.
\end{align} 
We denote the matrix $\phi^{-}$ to be
\begin{equation}
\phi^{-}=
\begin{pmatrix}
\phi^{-}_{11} & \phi^{-}_{12} & \phi^{-}_{13} & \phi^{-}_{14} & \phi^{-}_{15} \\
\phi^{-}_{21} & \phi^{-}_{22} & \phi^{-}_{23} & \phi^{-}_{24} & \phi^{-}_{25} \\
\phi^{-}_{31} & \phi^{-}_{32} & \phi^{-}_{33} & \phi^{-}_{34} & \phi^{-}_{35} \\
\phi^{-}_{41} & \phi^{-}_{42} & \phi^{-}_{43} & \phi^{-}_{44} & \phi^{-}_{45} \\
\phi^{-}_{51} & \phi^{-}_{52} & \phi^{-}_{53} & \phi^{-}_{54} & \phi^{-}_{55}
\end{pmatrix},
\end{equation}
and $\phi^{+}$ is denoted similarly.
Recall that $\alpha < 0$. If $Im(\lambda)>0$  
and $y<x$ then, $Re(e^{-i \lambda \alpha (x-y)})$ decays exponentially and so each integral of the first column of $\phi^{-}$ converges.
As a result, the components of the first column of $\phi^{-}$, are analytic in the upper half complex plane for 
$\lambda \in \mathbb{C}_{+}$, and 
continuous for $\lambda \in \mathbb{C}_{+} \cup \mathbb{R}$. 
\newline
In the same way for $y > x$, the components of the last four 
columns of $\phi^{+}$ are analytic in the upper half plane for 
$\lambda \in \mathbb{C}_{+}$ and continuous for 
$\lambda \in \mathbb{C}_{+} \cup \mathbb{R}$.
\\[3mm]
It is worth mentioning the case when $Im(\lambda) < 0$, then
the first column $\phi^{+}$ is analytic in the lower half plane for $\lambda \in \mathbb{C}_{-}$ and continuous for 
$\lambda \in \mathbb{C}_{-} \cup \mathbb{R}$, and the components of the last four columns of $\phi^{-}$ are analytic in the lower half plane for $\lambda \in \mathbb{C}_{-}$ and continuous for $\lambda \in \mathbb{C}_{-} \cup \mathbb{R}$.
\\[3mm]
Now, let us construct the Riemann-Hilbert problems.
To construct the Jost matrix in the upper-half plane we
note that \begin{equation}\label{psiphiplusminus-SASA}
\phi^{\pm} = \psi^{\pm} e^{-i \lambda \Lambda x}.
\end{equation}
Let $\phi^{\pm}_{j}$ be the $j$th column of $\phi^{\pm}$ for 
$j \in \{1,2,3,4,5\}$. Hence the first Jost matrix solution can be taken as
\begin{equation}\label{Pplusequ-SASA}
P^{(+)}(x,\lambda)=
(\phi_{1}^{-},\phi_{2}^{+},\phi_{3}^{+},\phi_{4}^{+},\phi_{5}^{+})
=\phi^{-} H_{1} + \phi^{+} H_{2},
\end{equation}
where $H_{1}=diag(1,0,0,0,0)$ and $H_{2}=diag(0,1,1,1,1)$.
\\[3mm]
Therefore $P^{(+)}$ is analytic for $\lambda \in \mathbb{C}_{+}$ and continuous 
for $\lambda \in \mathbb{C}_{+} \cup \mathbb{R}$.
\\[3mm]
For the lower-half plane, we can construct $P^{(-)} \in \mathbb{C}_{-}$ which is the analytic counterpart of $P^{(+)} \in \mathbb{C}_{+}$.
We do this by utilizing the equivalent spectral adjoint equation (\ref{adjointspatialphiinv-SASA}).
Because $\tilde{\phi}^{\pm}=(\phi^{\pm})^{-1}$ and  
$\psi^{\pm} = \phi^{\pm} e^{i \lambda \Lambda x}$, we have
\begin{equation}
(\phi^{\pm})^{-1} = e^{i \lambda \Lambda x} (\psi^{\pm})^{-1}.
\end{equation}
Let $\tilde{\phi}_{j}^{\pm}$ be the $j$th row of $\tilde{\phi}^{\pm}$
for $j \in \{1,2,3,4,5\}$. As above, we can get
\begin{equation}\label{Pminusequ-SASA}
P^{(-)}(x,\lambda)=\bigg( \tilde{\phi}_{1}^{-},\tilde{\phi}_{2}^{+},\tilde{\phi}_{3}^{+},
\tilde{\phi}_{4}^{+},\tilde{\phi}_{5}^{+}  
\bigg)^{T}=
H_{1}(\phi^{-})^{-1}+H_{2}(\phi^{+})^{-1}.
\end{equation}
Hence, $P^{(-)}$ is analytic for $\lambda \in \mathbb{C}_{-}$ and continuous 
for $\lambda \in \mathbb{C}_{-} \cup \mathbb{R}$. 
\\[3mm]
Since both $\phi^{-}$ and $\phi^{+}$ satisfy
\begin{equation}\label{phiTinvequation-SASA}
\phi^{T}(x,t,-\lam{}{}) =
C_{1} \phi^{-1}(x,t,\lam{}{}) C_{1}^{-1}.
\end{equation}
using (\ref{Pplusequ-SASA}), we have
\begin{equation}
P^{(+)}(x,t,-\lam{}{}) = 
\phi^{-}(x,t,-\lam{}{}) H_{1} + \phi^{+}(x,t,-\lam{}{}) H_{2}
\end{equation}
or equivalently
\begin{equation}\label{Pplustranspose-SASA}
(P^{(+)})^{T}(x,t,-\lam{}{}) = H_{1} (\phi^{-})^{T}(x,t,-\lam{}{}) 
+ H_{2} (\phi^{+})^{T}(x,t,-\lam{}{}).
\end{equation}
Substituting (\ref{phiTinvequation-SASA}) in (\ref{Pplustranspose-SASA}), we have the nonlocal 
symmetry property
\begin{equation}\label{PplusPminusrelation-SASA}
(P^{(+)})^{T}(x,t,-\lam{}{}) = 
C_{1} P^{(-)} (x,t,\lambda) C_{1}^{-1}.
\end{equation}
One can prove as well that
\begin{align}\label{PplusPplusrelation-SASA}
(P^{(+)})^{\dagger}(-x,-t,-\lambar{}{}) = 
C_{2} P^{(-)} (x,t,\lambda) C_{2}^{-1},
\\
\Pbar{}{(+)}(-x,-t,\lambar{}{}) = 
C_{3} P^{(+)} (x,t,\lambda) C_{3}^{-1}.
\end{align}
Employing analyticity of both $P^{(+)}$ and $P^{(-)}$, one can construct the Riemann-Hilbert problems 
\begin{equation}
P^{(-)}P^{(+)}=J,    
\end{equation}
where $J=e^{i \lambda \Lambda x} (H_{1}+H_{2}S)(H_{1}+S^{-1}H_{2}) e^{-i \lambda \Lambda x}$ for\quad $\lambda \in \mathbb{R}$.
\\[3mm]
Replacing (\ref{psiplusminus-SASA}) in (\ref{Pplusequ-SASA}), we have
\begin{equation}\label{Pplussimplified-SASA}
P^{(+)}(x,\lambda) = \phi^{+} (e^{i \lambda \Lambda x} S 
e^{-i \lambda \Lambda x} H_{1} +H_{2}).
\end{equation}
Because $\phi^{+}(x,\lambda) \rightarrow I_{5}$ when $x \rightarrow + \infty$, we get
\begin{equation}
\lim_{x \rightarrow + \infty} P^{(+)} = 
diag \big(s_{11}(\lambda),I_{4} \big)
\quad
\text{for}
\quad \lambda \in \mathbb{C}_{+} \cup \mathbb{R}.
\end{equation}
In the same way, 
\begin{equation}
\lim_{x \rightarrow - \infty} P^{(-)} = 
diag \big(\hat{s}_{11}(\lambda),I_{4} \big)
\quad
\text{for}
\quad \lambda \in \mathbb{C}_{-} \cup \mathbb{R}.
\end{equation}
Thus, if we choose
\begin{align}
\label{GplusPplusrelation-SASA}
G^{(+)}(x,\lambda) &= P^{(+)}(x,\lambda) 
diag \big(s_{11}^{-1} (\lambda),I_{4} \big),
\\ \label{GminusPminusrelation-SASA}
(G^{(-)})^{-1}(x,\lambda) &= 
diag \big(\hat{s}_{11}^{-1} (\lambda),I_{4} \big)
P^{(-)}(x,\lambda),
\end{align}
the two generalized matrices
$G^{(+)}(x,\lambda)$ and $G^{(-)}(x,\lambda)$ generate the matrix Riemann-Hilbert problems on the real line for the resulting nonlocal Sasa-Satsuma equation, given by
\begin{equation}\label{GplusGminusG0-SASA}
G^{(+)}(x,\lambda) = G^{(-)}(x,\lambda) G_0(x,\lambda),
\quad
\text{for}
\quad \lambda \in \mathbb{R},
\end{equation}
where the jump matrix $G_0(x,\lambda)$ can be cast as
\begin{equation}\label{Gequ-SASA}
G_0(x,\lambda) = 
diag \big(\hat{s}_{11}^{-1} (\lambda),I_{4} \big) \,
J \,
diag \big(s_{11}^{-1} (\lambda),I_{4} \big),
\end{equation}
which reads
\begin{equation}
G_0^{}(x,\lambda) =
\begin{pmatrix}
s_{11}^{-1} \hat{s}_{11}^{-1} 
& 
\hat{s}_{12}\hat{s}_{11}^{-1} e^{i \lambda \alpha x}
& 
\hat{s}_{13}\hat{s}_{11}^{-1} e^{i \lambda \alpha x}
& 
\hat{s}_{14}\hat{s}_{11}^{-1} e^{i \lambda \alpha x}
& 
\hat{s}_{15}\hat{s}_{11}^{-1} e^{i \lambda \alpha x}
\\[2mm]
s_{21} s_{11}^{-1} e^{-i \lambda \alpha x} 
& 
1
& 
0
& 
0
&
0
\\[4mm]
s_{31} s_{11}^{-1} e^{-i \lambda \alpha x}
& 
0
& 
1
&
0
&
0
\\[4mm]
s_{41} s_{11}^{-1} e^{-i \lambda \alpha x}
& 
0 
& 
0
& 
1
&
0
\\[4mm]
s_{51} s_{11}^{-1} e^{-i \lambda \alpha x}
& 
0 
& 
0
& 
0
& 
1
\end{pmatrix},
\end{equation}
and its canonical normalization conditions:
\begin{align}
G^{(+)}(x,\lambda) \rightarrow I_{5} \quad \text{as} \quad \lambda \in \mathbb{C}_{+} \cup \mathbb{R} \rightarrow \infty,
\\
G^{(-)}(x,\lambda) \rightarrow I_{5} \quad \text{as} \quad \lambda \in \mathbb{C}_{-} \cup \mathbb{R} \rightarrow \infty.
\end{align}
From (\ref{PplusPminusrelation-SASA}) along with (\ref{GplusPplusrelation-SASA})-(\ref{GminusPminusrelation-SASA}) and using (\ref{s11hats11relation-SASA})-(\ref{s11hats11relationC3-SASA}), we deduce the nonlocal involution properties
\begin{equation}\label{GplusdaggerGminusinv-SASA}
\begin{cases}
(G^{(+)})^{T}(x,t,-\lam{}{}) = C_{1} (G^{(-)})^{-1} (x,t,\lam{}{}) C_{1}^{-1},
\\
(G^{(+)})^{\dagger}(-x,-t,-\lambar{}{}) = C_{2} (G^{(-)})^{-1} (x,t,\lambda) C_{2}^{-1},
\\
\Gmatbar{}{(+)}(-x,-t,\lambar{}{}) = C_{3} G^{(+)}(x,t,\lam{}{}) C_{3}^{-1}.
\end{cases}
\end{equation}
Furthermore, from (\ref{GplusGminusG0-SASA}), (\ref{Gequ-SASA}),(\ref{GplusdaggerGminusinv-SASA}),
and (\ref{s11hats11relation-SASA})-(\ref{s11hats11relationC3-SASA}), we derive the following nonlocal involution properties for $G_0$
\begin{equation}
\begin{cases}
G_0^{T}(x,t,-\lam{}{}) = C_{1} G_0(x,t,\lam{}{}) C_{1}^{-1},
\\
G_0^{\dagger}(-x,-t,-\lambar{}{}) = C_{2} G_0(x,t,\lam{}{}) C_{2}^{-1},
\\
\Gmatbar{0}{}(-x,-t,\lambar{}{}) = C_{3} G_0(x,t,\lam{}{}) C_{3}^{-1},
\end{cases}
\lam{}{} \in \mathbb{R}.
\end{equation}
\subsection{Time evolution of the scattering data}
\noindent
At this point, we have to determine how the scattering data evolves over time. In order to do that, we differentiate equation (\ref{psiplusminus-SASA}) with respect to time $t$ and applying (\ref{temporalphi-SASA}) gives 
\begin{equation}
S_{t} =i \lam{}{5} [\Omega,S],
\end{equation}
and thus
\begin{equation}
S_{t}=
\begin{pmatrix}
0 & i \beta \lam{}{5} s_{12} & 
i \beta \lam{}{5} s_{13} & i \beta \lam{}{5} s_{14} & \beta \lam{}{5} s_{15} 
\\
-i \beta \lam{}{5} s_{21} & 0 & 
0 & 0 & 0
\\
-i \beta \lam{}{5} s_{31} & 0 & 
0 & 0 & 0
\\
-i \beta \lam{}{5} s_{41} & 0 & 
0 & 0 & 0
\\
-i \beta \lam{}{5} s_{51} & 0 & 
0 & 0 & 0
\end{pmatrix}.
\end{equation}
As a result, we have
\begin{equation}
\begin{cases}
\begin{aligned}[c]
s_{12}(t,\lambda) = s_{12}(0,\lambda) e^{i \beta \lam{}{3} t},
\\
s_{13}(t,\lambda) = s_{13}(0,\lambda) e^{i \beta \lam{}{3} t},
\\
s_{14}(t,\lambda) = s_{14}(0,\lambda) e^{i \beta \lam{}{3} t},
\\
s_{15}(t,\lambda) = s_{15}(0,\lambda) e^{i \beta \lam{}{3} t},
\end{aligned}
\quad
\begin{aligned}[c]
s_{21}(t,\lambda) = s_{21}(0,\lambda) e^{-i \beta \lam{}{3} t},
\\
s_{31}(t,\lambda) = s_{31}(0,\lambda) e^{-i \beta \lam{}{3} t}, 
\\
s_{41}(t,\lambda) = s_{41}(0,\lambda) e^{-i \beta \lam{}{3} t},
\\
s_{51}(t,\lambda) = s_{51}(0,\lambda) e^{-i \beta \lam{}{3} t},
\end{aligned}
\end{cases}
\end{equation}
and $s_{11},s_{2i},s_{3i},s_{4i},s_{5i}$ 
are constants for $i \in \{2,...,5\}$.
\section{Soliton solutions}
\subsection{General case}
\noindent
The determinant of the matrix $G^{(\pm)}$ determines the type of soliton solutions generated using the Riemann-Hilbert problems. In the regular case, when
$det(G^{(\pm)}) \neq 0$, we obtain the unique solution. In the non-regular case, that is to say when $det(G^{(\pm)})=0$, it could generate discrete eigenvalues in the spectral plane. This non-regular case can be transformed into the regular case to solve for soliton solutions \cite{Yang2010}.
\newline
From (\ref{Pplussimplified-SASA}) and $det(\phi^{\pm})=1$, we can show that
\begin{align}\label{ppluss11-SASA}
det(P^{(+)}(x,\lambda))&=s_{11}(\lambda),
\\ \label{pminuss11hat-SASA}
det(P^{(-)}(x,\lambda))&=\hat{s}_{11}(\lambda).
\end{align}
Because $det(S(\lambda))=1$, this implies that
$S^{-1}(\lambda)=\bigg( cof(S(\lambda)) \bigg)^{T}$, and
\begin{equation}
\hat{s}_{11}=
\begin{vmatrix}
s_{22} & s_{23} & s_{24} & s_{25} \\
s_{32} & s_{33} & s_{34} & s_{35} \\
s_{42} & s_{43} & s_{44} & s_{45} \\
s_{52} & s_{53} & s_{54} & s_{55} 
\end{vmatrix},
\end{equation}
which should be zero for the non-regular case.
\newline
In order to generate soliton solutions, we need the solutions of
$det(P^{(+)}(x,\lambda))=det(P^{(-)}(x,\lambda))=0$ to be simple.
When
$det(P^{(+)}(x,\lambda))=s_{11}(\lambda)=0$, we assume
$s_{11}(\lambda)$ has simple zeros with
discrete eigenvalues $\lambda_{k} \in \mathbb{C}_{+}$ for 
$k \in \{1,2,...,2N_{1}=N\}$, while for $det(P^{(-)}(x,\lambda))=\hat{s}_{11}(\lambda)=0$, 
we assume
$\hat{s}_{11}(\lambda)$ has simple zeros with
discrete eigenvalues $\hat{\lambda}_{k} \in \mathbb{C}_{-}$ for 
$k \in \{1,2,...,2N_{1}=N\}$, which are the poles of the transmission coefficients \cite{DrazinJohnson1989}.
\newline
From equations (\ref{s11hats11relation-SASA})-(\ref{s11hats11relationC3-SASA})
and $det(P^{(\pm)}(x,\lambda))=0$,
one can deduce that 
\begin{equation}\label{lambdacondition-SASA}
\text{if} \quad \lam{}{} \in \mathbb{C}_{+},
\quad
\text{then} \quad
\begin{cases}
- \lam{}{} \in \mathbb{C}_{-},
\\
- \lambar{}{} \in \mathbb{C}_{+},
\\
\lambar{}{} \in \mathbb{C}_{-},
\end{cases}
\lam{}{} \notin \i \mathbb{R}.
\end{equation}
If $\lam{}{} = \i m \in \i \mathbb{R}$, for $m > 0$,
the couple ($\lam{}{},-\lambar{}{}) \in \mathbb{C}_{+}^{2}$ coincide, forcing
$\lamhat{}{}=-\lam{}{} =-i m \in \mathbb{C}_{-}$.
\noindent
\newline
To make this more clear, we can view the choices of the eigenvalues in a more systematic way. Recall that the Riemann-Hilbert problem requires the same number of eigenvalues in the upper-half plane and in the lower-half plane. Assume $\lam{k}{} \in \mathbb{C}_{+}$ 
for all $k=1,2,\ldots,2N_{1}$. 
Fix $n$ for $1 \leq n \leq N_{1}$ and let $\lam{n}{}$ lies off the imaginary axis. The eigenvalues in the upper-half plane are given by the $N_{1}$-couples $(\lam{n}{},\lam{N_{1}+n}{}) = (\lam{}{},-\lambar{}{}) \in \mathbb{C}_{+}^{2}$,
which are assumed to be the zeros of $det(P^{(+)}(x,\lambda))=0$. For any $\lam{n}{}$, the choice of $\lam{N_{1}+n}{}$ depends on $\lam{n}{}$, that is, $\lam{n}{} = - \lambar{N_{1}+n}{}$, where
$\lam{n}{}$ is freely chosen. If $\lam{n}{}$ lies on the imaginary axis, then the pair of eigenvalues coincide.
\\[3mm]
In the lower-half plane, the eigenvalues are determined by the choice of the eigenvalue $\lam{n}{}$ in the upper-half plane. We have $\lamhat{k}{} \in \mathbb{C}_{-}$ 
for all $k=1,2,\ldots,2N_{1}$ and similarly the eigenvalues are given by the $N_{1}$-couples $(\lamhat{n}{},\lamhat{N_{1}+n}{}) =
(-\lam{}{}, \lambar{}{}) \in \mathbb{C}_{-}^{2}$, which are assumed to be the zeros of $det(P^{(-)}(x,\lambda))=0$,
and $\lamhat{n}{} = - \lambarhat{N_{1}+n}{}$.
\\
In other words, if $\lam{n}{}$ is not pure imaginary, then the scheme of the eigenvalues take the form
\begin{equation}\label{EigenvaluesConfiguration-SASA}
(\lam{n}{},\lam{N_{1}+n}{},\lamhat{n}{},\lamhat{N_{1}+n}{})
=(\lam{}{},-\lambar{}{},-\lam{}{},\lambar{}{}).  
\end{equation}
Each $Ker(P^{(+)}(x,\lambda_{k}))$
contains only a single column vector $\vvec{k}{}$, similarly each $Ker(P^{(-)}(x,\hat{\lambda}_{k}))$ contains only a single row vector $\vvechat{k}{}$ such that:
\begin{equation}\label{Pplus-SASA}
P^{(+)}(x,\lambda_{k}) \vvec{k}{}=0 \quad \text{for} \quad k \in \{1,2,...,2N_{1}\},
\end{equation}
and
\begin{equation}\label{Pminus-SASA}
\vvechat{k}{} P^{(-)}(x,\hat{\lambda}_{k})=0 \quad \text{for} \quad k \in \{1,2,...,2N_{1}\}.
\end{equation}
To obtain explicit soliton solutions, we take $G_0=I_{5}$ in the
Riemann-Hilbert problems. This will force the reflection coefficients 
$s_{21}=s_{31}=s_{41}=s_{51}=0$ and $\hat{s}_{12}=\hat{s}_{13}=\hat{s}_{14}=\hat{s}_{15}=0$.
\newline
In that case, the Riemann-Hilbert problems can be presented as follows \cite{MaAugust2020}: 
\begin{equation}\label{Pplussum-SASA}
G^{(+)}(x,\lambda)=I_{5}-\sum\limits_{k,j=1}^{N} 
\frac{\vvec{k}{}(M^{-1})_{kj}\vvechat{j}{}}{\lambda-\hat{\lambda}_{j}},
\end{equation}
and
\begin{equation}\label{Pminussum-SASA}
(G^{(-)})^{-1}(x,\lambda)=I_{5}+\sum\limits_{k,j=1}^{N} 
\frac{\vvec{k}{}(M^{-1})_{kj}\vvechat{j}{}}{\lambda-\lambda_{k}},
\end{equation}
where $M=(m_{kj})_{N \times N}$ is a matrix defined by \cite{MaAugust2020}
\begin{equation}\label{Mmatrixcondition-SASA}
m_{kj}=
\begin{cases}
\frac{\vvechat{k}{} \vvec{j}{}}{\lambda_{j}-\hat{\lambda}_{k}},
& \text{if} \quad
\lambda_{j} \neq \hat{\lambda}_{k},
\\
0, & \text{if} \quad
\lambda_{j} = \hat{\lambda}_{k}.
\end{cases}
\quad
k,j \in \{1,2,...,N\},
\end{equation}
Since the zeros $\lambda_{k}$ and $\hat{\lambda}_{k}$ are constants, because they are independent of space and time, 
we can explore the spatial and temporal evolution of the scattering vectors $\vvec{k}{}(x,t)$ and 
$\vvechat{k}{}(x,t)$, $1\leq k \leq N$.
\\[3mm]
Taking the $x$-derivative of both sides of the equation
\begin{equation}
P^{(+)}(x,\lambda_{k})\vvec{k}{}=0, \quad 1\leq k \leq N,
\end{equation}
and knowing that $P^{(+)}$ satisfies the spectral
spatial equivalent equation (\ref{spatialphi-SASA}) together with
(\ref{Pplus-SASA}), we obtain
\begin{equation}
P^{(+)}(x,\lambda_{k}) 
\Bigg( \frac{d\vvec{k}{}}{dx}-i\lambda_{k} \Lambda \vvec{k}{} \Bigg)=0
\quad
\text{for}
\quad
k,j \in \{1,2,...,N\}.
\end{equation}
In a similar manner, taking the $t$-derivative and using the temporal equation (\ref{temporalphi-SASA}) with (\ref{Pplus-SASA}), we acquire
\begin{equation}
P^{(+)}(x,\lam{k}{}) 
\Bigg( \frac{d\vvec{k}{}}{dt}-i \lam{k}{5} \Omega \vvec{k}{} \Bigg)=0
\quad
\text{for}
\quad
k,j \in \{1,2,...,N\}.
\end{equation}
For the adjoint spectral equations (\ref{adjointspatialphi-SASA}) and (\ref{adjointtemporalphi-SASA}), we can obtain the following similar results
\begin{equation}
\Bigg( \frac{d\vvechat{k}{}}{dx}+i\hat{\lambda}_{k} \vvechat{k}{} \Lambda \Bigg) P^{(-)}(x,\hat{\lam{}{}}_{k})=0,
\end{equation}
and
\begin{equation}
\Bigg( \frac{d\vvechat{k}{}}{dt}+i\hat{\lambda}^{5}_{k} \vvechat{k}{} \Omega \Bigg) P^{(-)}(x,\hat{\lam{}{}}_{k})=0.
\end{equation}
Because $\vvec{k}{}$ is a single vector in the kernel of $P^{(+)}$,
so
$\frac{d \vvec{k}{}}{dx} -\i \lambda_{k} \Lambda \vvec{k}{}$
and 
$\frac{d \vvec{k}{}}{dt} -\i \lambda^{5}_{k} \Omega \vvec{k}{}$
\\
are scalar multiples of $\vvec{k}{}$. 
\newline 
Hence without loss of generality we can take the space dependence of $\vvec{k}{}$ to be:
\begin{equation}\label{vkxderivative-SASA}
\frac{d \vvec{k}{}}{dx}=i\lambda_{k} \Lambda \vvec{k}{}, \quad 1\leq k \leq N
\end{equation}
and the time dependence of $v_{k}$ as:
\begin{equation}\label{vktderivative-SASA}
\frac{d \vvec{k}{}}{dt}=i \lambda^{5}_{k} \Omega \vvec{k}{}, \quad 1\leq k \leq N.
\end{equation}
so, we can conclude that
\begin{equation}\label{vkequation-SASA}
\vvec{k}{}(x,t)=\vvec{k}{}(x,t,\lambda_{k})=e^{i\lambda_{k}\Lambda x + i\lambda^{5}_{k} \Omega t} \wvec{k}{}
\quad \text{for} \quad k \in \{1,2,...,N\},
\end{equation}
by solving equations (\ref{vkxderivative-SASA}) and (\ref{vktderivative-SASA}). Likewise, we get
\begin{equation}\label{vkhatequation-SASA}
\vvechat{k}{}(x,t)=\vvechat{k}{}(x,t,\hat{\lambda}_{k})=
\wvechat{k}{} e^{-i\hat{\lambda}_{k}\Lambda x - i\hat{\lambda}^{5}_{k} \Omega t} 
\quad \text{for} \quad k \in \{1,2,...,N\},
\end{equation}
where $\wvec{k}{}$ and $\wvechat{k}{}$ are constant column and row vectors in $\mathbb{C}^{5}$, respectively. In addition, they need to satisfy the orthogonality condition:
\begin{equation}\label{wkwlorthogonality-SASA}
\wvechat{k}{} \wvec{l}{} = 0, \quad \text{when} \quad \lam{l}{}=\lamhat{k}{}, \quad 1 \leq k,l \leq N.
\end{equation}
From (\ref{Pplus-SASA}) and using the formula (\ref{PplusPminusrelation-SASA}),
it is easy to see 
\begin{equation}
\vvec{k}{T}(x,t,-\lam{k}{}) (P^{(+)})^{T}(x,t,-\lam{k}{})=
\vvec{k}{T}(x,t,-\lam{k}{}) C_{1} P^{(-)}(x,t,\lam{k}{}) C_{1}^{-1} = 0.
\end{equation}
Because $\vvec{k}{T}(x,t,-\lam{k}{}) C_{1} P^{-}(x,t,\lam{k}{})$
can be zero and using (\ref{Pminus-SASA}) this leads to
\begin{align}
\vvec{k}{T}(x,t,-\lam{k}{}) C_{1} P^{(-)}(x,t,\lam{k}{}) 
&=
\vvec{k}{T}(x,t,\lam{k}{}) C_{1} P^{(-)}(x,t,-\lam{k}{})
=0
\\
&=
\vvechat{k}{}(x,t,\hat{\lambda}_{k}) P^{(-)}(x,t,\hat{\lambda}_{k}),
\end{align}
thus, we can take
\begin{equation}\label{vdaggerC1-SASA}
\vvechat{k}{}(x,t,\hat{\lambda}_{k})=
\vvec{k}{T}(x,t,\lam{k}{}) C_{1}.
\end{equation}
Therefore, the involution relations (\ref{vkequation-SASA}) and (\ref{vkhatequation-SASA}) give
\begin{equation}\label{vknon-localequ-SASA}
\vvec{k}{}(x,t)=e^{i\lambda_{k} \Lambda x 
+ i\lam{k}{5} \Omega t} \wvec{k}{},
\end{equation}
\begin{equation}\label{vkhatnon-localequ-SASA}
\vvechat{k}{}(x,t)= \wvec{k}{T} e^{-i\hat{\lambda}_{k} \Lambda x 
- i\hat{\lambda}_{k}^{5} \Omega t} C_{1}.
\end{equation}
Now, in order to satisfy the orthogonality condition (\ref{wkwlorthogonality-SASA}), one can notice that we require:
\begin{equation}\label{Orthogonality1-SASA}
\wvec{k}{\dagger} C_{2} \wvec{l}{} = 0, \quad \text{as} \quad \lam{l}{}=\lamhat{k}{}, \quad 1 \leq k,l \leq N.
\end{equation}
due to the fact that, $\lamhat{k}{}=-\lambar{k}{}=\lam{k}{}$ still occurs for 
$\lam{k}{} \in \i \mathbb{R}$, while $\lamhat{k}{}=-\lambar{k}{}$
holds, when $\lam{k}{} \neq \lamhat{k}{}$.
\newline
Because the jump matrix $G_{0}=I_{5}$, we can solve the Riemann-Hilbert problem precisely. As a result, we can determine the potentials by computing the matrix $P^{(+)}$. Because $P^{(+)}$ is analytic, we can expand $G^{(+)}$ as follows:
\begin{equation}\label{Gexpansion-SASA}
G^{(+)}(x,\lambda)=I_{5}+\frac{1}{\lambda} G^{(+)}_{1}(x)+
O \Big( \frac{1}{\lambda^{2}} \Big),
\quad
\text{when}
\quad
\lambda \rightarrow \infty.
\end{equation}
Knowing that $G^{(+)}$ satisfies the spectral problem, substituting
it in (\ref{spatialphi-SASA}) and matching the coefficients of the same power of $\frac{1}{\lambda}$, at order $O(1)$,
we get
\begin{equation}\label{G1plusequs-SASA}
P=-[\Lambda,G^{(+)}_{1}].
\end{equation}
If we denote 
\begin{equation}
G_{1}^{(+)}=
\begin{pmatrix}
\Gsol{11}{} & \Gsol{12}{} & \Gsol{13}{} & \Gsol{14}{} & \Gsol{15}{}  \\
\Gsol{21}{} & \Gsol{22}{} & \Gsol{23}{} & \Gsol{24}{} & \Gsol{25}{} \\
\Gsol{31}{} & \Gsol{32}{} & \Gsol{33}{} & \Gsol{34}{} & \Gsol{35}{} \\
\Gsol{41}{} & \Gsol{42}{} & \Gsol{43}{} & \Gsol{44}{} & \Gsol{45}{} \\
\Gsol{51}{} & \Gsol{52}{} & \Gsol{53}{} & \Gsol{54}{} & \Gsol{55}{} \\
\end{pmatrix}
\end{equation}
then, since $P$ satisfies the symmetry relations (\ref{Pequ-SASA}) simultaneously,
therefore from (\ref{G1plusequs-SASA}), $G_{1}^{(+)}$ satisfies the following symmetry relations:
\begin{align}\label{G1plussymmetryrelationC0}
(\Gmat{1}{(+)})^{\dagger}(x,t) &= \Gmat{1}{(+)}(x,t)
\\
(\Gmat{1}{(+)})^{T}(-x,-t) &= C_{1} \Gmat{1}{(+)}(x,t)
C_{1}^{-1}
\\
(\Gmat{1}{(+)})^{\dagger}(-x,-t) &= C_{2} \Gmat{1}{(+)}(x,t)
C_{2}^{-1}
\\ \label{G1plussymmetryrelationC3}
\Gmatbar{1}{(+)}(-x,-t) &= C_{3} \Gmat{1}{(+)}(x,t)
C_{3}^{-1}
\end{align}
accordingly, it can be reduced to the form
\begin{equation}
G_{1}^{(+)}=
\begin{pmatrix}
\Gsol{11}{}(x,t) & \Gsol{12}{}(x,t) & \Gsol{12}{}(-x,-t) & \Gbarsol{12}{}(x,t) & \Gbarsol{12}{}(-x,-t)  \\
\Gbarsol{12}{}(x,t) & \Gsol{22}{}(x,t) & \Gbarsol{32}{}(x,t) & \Gbarsol{42}{}(x,t) & \Gbarsol{52}{}(-x,-t) \\
\Gbarsol{12}{}(-x,-t) & \Gsol{32}{}(x,t) & \Gbarsol{22}{}(-x,-t) & \Gbarsol{52}{}(-x,-t) & \Gbarsol{42}{}(-x,-t) \\
\Gsol{12}{}(x,t) & \Gsol{42}{}(x,t) & \Gsol{52}{}(-x,-t) & \Gsol{22}{}(x,t) & \Gbarsol{32}{}(-x,-t) \\
\Gsol{12}{}(-x,-t) & \Gsol{52}{}(x,t) & \Gsol{42}{}(-x,-t) & \Gsol{32}{}(-x,-t) & \Gbarsol{22}{}(-x,-t) \\
\end{pmatrix}
\end{equation}
thus,
\begin{equation}\label{PG1commutator-SASA}
P=-[\Lambda,G_{1}^{(+)}]=
\begin{pmatrix}
0 & 
-\alpha \Gsol{12}{}(x,t) & 
-\alpha \Gsol{12}{}(-x,-t) & 
-\alpha \Gbarsol{12}{}(x,t) & 
-\alpha \Gbarsol{12}{}(-x,-t)   \\
\alpha \Gbarsol{12}{}(x,t) & 0 & 0 & 0 & 0 \\
\alpha \Gbarsol{12}{}(-x,-t) & 0 & 0 & 0 & 0 \\
\alpha \Gsol{12}{}(x,t) & 0 & 0 & 0 & 0 \\
\alpha \Gsol{12}{}(-x,-t) & 0 & 0 & 0 & 0 
\end{pmatrix}.
\end{equation}
Matching the components of (\ref{PG1commutator-SASA}) to the components of the $P$ matrix, we can recover the potential 
$\u{}{}$  
\begin{align}\label{uvequations-SASA}
\u{}{}(x,t) &= -\alpha \Gsol{12}{}(x,t),
\end{align}
Note that all components in (\ref{PG1commutator-SASA}) are equivalent and compatible to the components in (\ref{Pmatrix-SASA}). It can be seen from (\ref{Gexpansion-SASA}) that
\begin{equation}
G_{1}^{(+)} = \lambda \lim_{\lambda \rightarrow \infty} (G^{(+)}(x,\lambda)-I_{5}),
\end{equation}
then using equation (\ref{Pplussum-SASA}), we deduce
\begin{equation}\label{G1plussummation-SASA}
G_{1}^{(+)} = - \sum\limits_{k,j=1}^{N} \vvec{k}{} (M^{-1})_{k,j} \vvechat{j}{},
\end{equation}
where 
\begin{equation}
\vvec{k}{}=(\vvec{k,1}{},\vvec{k,2}{},\vvec{k,3}{},\vvec{k,4}{},\vvec{k,5}{})^{T},\vvechat{k}{} =(\vvechat{k,1}{},\vvechat{k,2}{},\vvechat{k,3}{},\vvechat{k,4}{},\vvechat{k,5}{}).
\end{equation}
We deduce that the specific Riemann-Hilbert problem solutions determined by (\ref{Pplussum-SASA})-(\ref{Mmatrixcondition-SASA}), 
satisfy (\ref{GplusdaggerGminusinv-SASA}). Hence the matrix $G_{1}^{(+)}$ posses the symmetry relations (\ref{G1plussymmetryrelationC0})-(\ref{G1plussymmetryrelationC3}), which are generated from the non-local symmetry (\ref{Ureduction-SASA}).
\\
Now, by substituting (\ref{G1plussummation-SASA}) into (\ref{uvequations-SASA})
and using (\ref{vknon-localequ-SASA}) and (\ref{vkhatnon-localequ-SASA}), 
we generate the $N$-soliton solution to the nonlocal fifth-order Sasa-Satsuma equation
\begin{align}
\u{}{}(x,t) &= \alpha \sum\limits_{k,j=1}^{N} 
\vvec{k,1}{} (M^{-1})_{kj}
\vvechat{j,2}{}.
\end{align}
\section{Exact soliton solutions and dynamics}
\subsection{Explicit one-soliton solution}
\noindent
For a general explicit formula for the one-soliton solution of the Sasa-Satsuma equation 
(\ref{Sasa-Satsuma-equation-u-SASA}), i.e., when $N=1$, the eigenvalues configuration gives $\lambda_{1}=\i m$ and $\hat{\lambda}_{1}=-\i m$, where $m > 0$, in order to fulfill condition (\ref{lambdacondition-SASA}). 
Taking $\wvec{1}{}$ to be the vector
$\wvec{1}{}=(\wvec{11}{},\wvec{12}{},-\wvec{12}{},
\wvec{12}{},-\wvec{12}{})^{T}$,
for
$\wvec{11}{},\wvec{12}{} \in \mathbb{R} \backslash \{0\}$
The explicit solution yields:
\begin{align}\label{usolution-SASA}
\u{}{}(x,t)&= 
\frac{
\i 2 \alpha m \wvec{11}{} \wvec{12}{}
}{ \wvec{11}{2}
e^{-\alpha m x - \beta m^{5} t}
+4
\wvec{12}{2}
e^{\alpha m x + \beta m^{5} t}
}.
\end{align}
Since this Sasa-Satsuma equation require the orthogonality condition
\begin{equation}\label{Orthogonality2-SASA}
\wvec{1}{\dagger} C_{2} \wvec{1}{} = 0,
\end{equation}
resulting in $\wvec{11}{2}=4 \wvec{12}{2}$.
\newline
As a consequence, the solution to the scalar nonlocal reverse-spacetime Sasa-Satsuma equation (\ref{Sasa-Satsuma-coupled1-SASA})  in the one-soliton case simplifies to
\begin{align}\label{usolution-SASA-2}
\u{}{}(x,t)&= 
\pm \i \frac{1}{2} \alpha m  
sech (\alpha m x + \beta m^{5} t).
\end{align}
\subsubsection{\textbf{Dynamics of the one-soliton}}
\begin{figure}[H]
\begin{center}
\fcolorbox{gray}{black}{\includegraphics[width=0.45\textwidth,height=4.5cm]{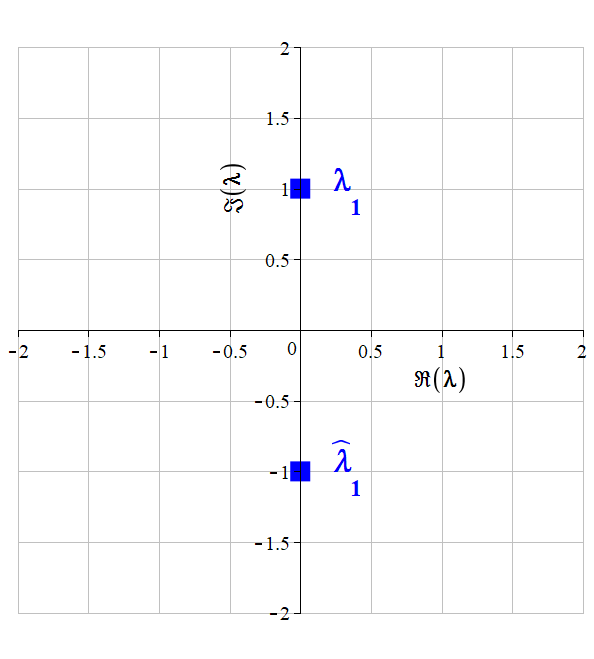}}
\fcolorbox{gray}{black}{\includegraphics[width=0.45\textwidth,height=4.5cm]{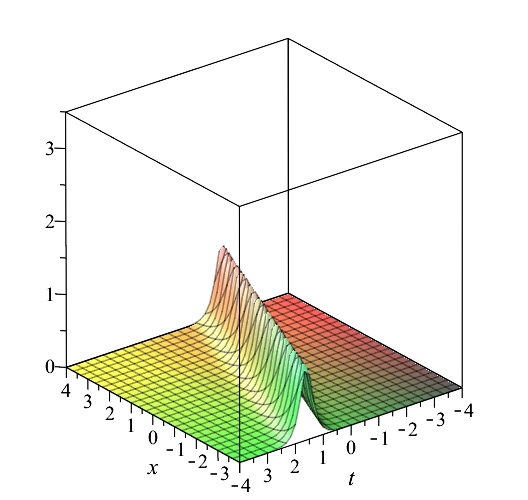}}
\\
\fcolorbox{gray}{black}{\includegraphics[width=0.45\textwidth,height=4.5cm]{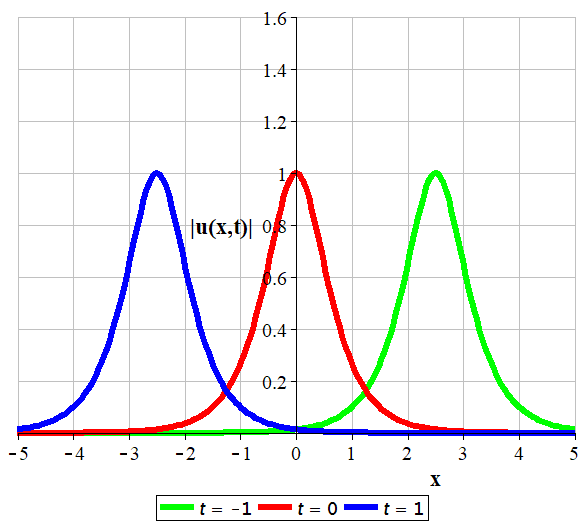}}
\fcolorbox{gray}{black}{\includegraphics[width=0.45\textwidth,height=4.5cm]{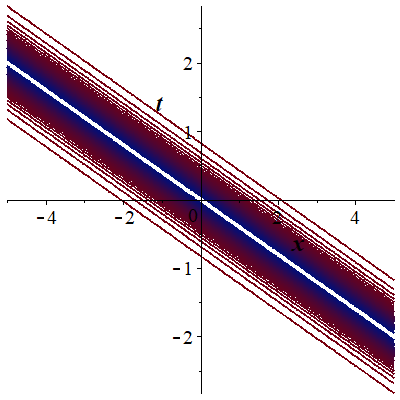}}
\caption{Spectral plane along with 3D, 2D and contours plots of $|u(x,t)|$ of the travelling one-soliton with parameters
$(\alpha,\beta)=(-2,-5)$, 
$(\lambda_{1},\hat{\lambda}_{1})=(i,-i)$, 
$\wvec{1}{}=(2,1,-1,1,-1)$.}%
\label{1solitonplot3-SASA}%
\end{center}
\end{figure}

\begin{figure}[H]
\begin{center}
\fcolorbox{gray}{black}{\includegraphics[width=0.45\textwidth,height=4.5cm]{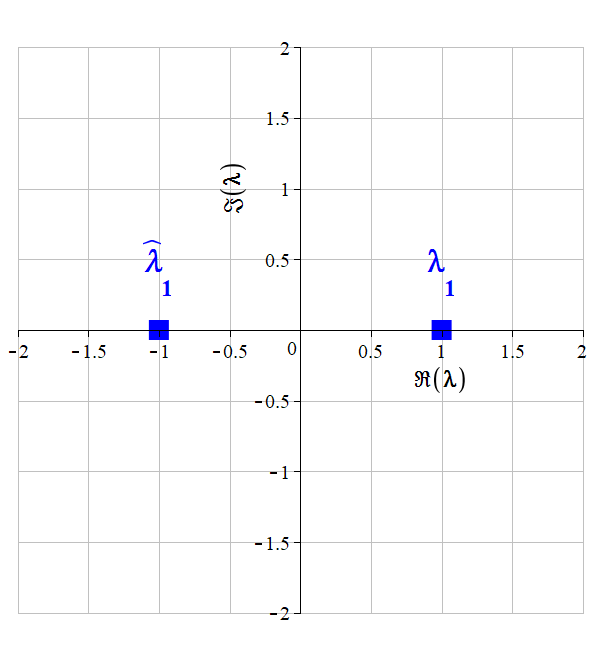}}
\fcolorbox{gray}{black}{\includegraphics[width=0.45\textwidth,height=4.5cm]{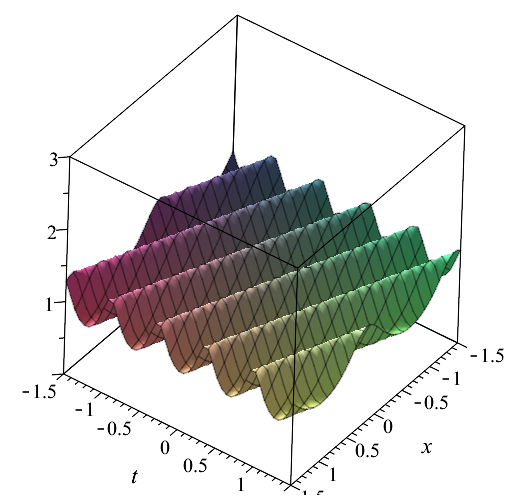}}
\\
\fcolorbox{gray}{black}{\includegraphics[width=0.45\textwidth,height=4.5cm]{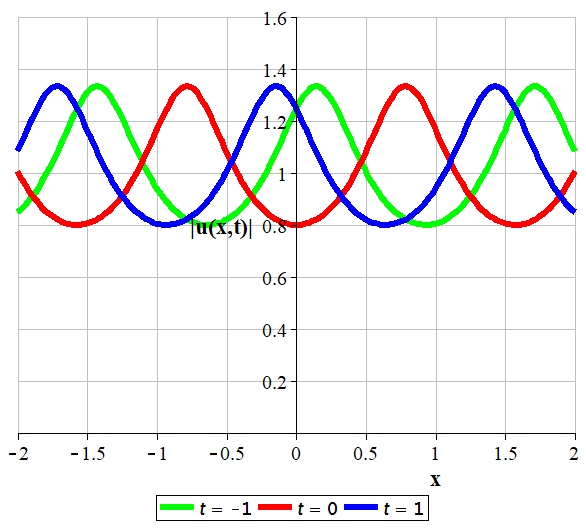}}
\fcolorbox{gray}{black}{\includegraphics[width=0.45\textwidth,height=4.5cm]{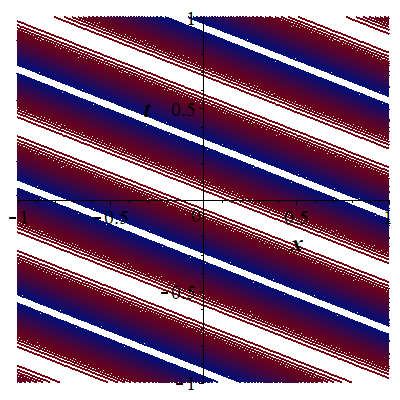}}
\caption{Spectral plane along with 3D, 2D and contours plots of $|u(x,t)|$ of the one-soliton breather with parameters  
$(\alpha,\beta)=(-2,-5)$, 
$(\lambda_{1},\hat{\lambda}_{1})=(-2,-5)$, 
$\wvec{1}{}=(1,1,1,1,1)$.}
\label{1solitonplot4-SASA}
\end{center}
\end{figure}

\noindent
For the one-soliton, the soliton moves
with speed 
$V=\frac{\beta}{\alpha} m^{4}$ 
along the line 
$x=\frac{\beta}{\alpha} m^{4} t$. In that case, the amplitude is given by
$|u(x,t)|= \frac{1}{2} \alpha m$.
The amplitude of the moving soliton stays constant as seen in figure \ref{1solitonplot3-SASA}. In the case when $\lam{1}{}=m$ is real, we get a breather with period $\frac{\pi}{|\beta m^{5}|}$ as in figure \ref{1solitonplot4-SASA}.
\newpage
\vspace{-1cm}
\subsection{Explicit two-soliton solutions}
\begin{figure}[H]
\centering
\begin{subfigure}{0.30\textwidth}
\includegraphics[width=5cm,height=5cm]{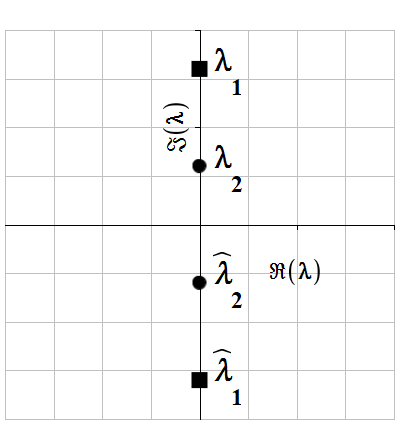}
\caption{Case I}
\label{fig:first-SASA}
\end{subfigure}
\begin{subfigure}{0.30\textwidth}
\includegraphics[width=5cm,height=5cm]{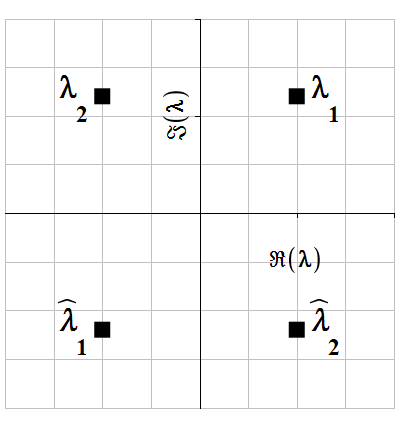}
\caption{Case II}
\label{fig:second-SASA}
\end{subfigure}
\begin{subfigure}{0.30\textwidth}
\includegraphics[width=5cm,height=5cm]{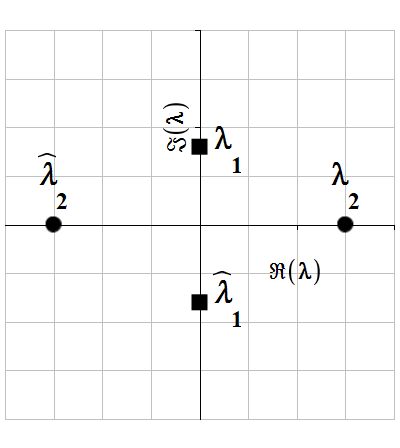}
\caption{Case III}
\label{fig:third-SASA}
\end{subfigure}
\caption{Spectral planes of two-soliton eigenvalues cases}
\label{2-soliton-Sasa-Satsuma-cases-SASA}
\end{figure}

\noindent
For a general explicit formula for two-soliton solutions of the Sasa-Satsuma equation 
(\ref{Sasa-Satsuma-equation-u-SASA}), i.e., when $N=2$, the configuration of the eigenvalues for this equation is given by
$(\lam{1}{},\lam{2}{},\lamhat{1}{},\lamhat{2}{})
=(\lam{}{},-\lambar{}{},-\lam{}{},\lambar{}{})$.
As a result, we have three distinct cases as shown in figure \ref{2-soliton-Sasa-Satsuma-cases-SASA}. 
In all cases, the eigenvalues
$\lam{1}{}, \lam{2}{} \in \mathbb{C}_{+} \cup \mathbb{R}$ and 
$\hat{\lambda}_{1}, \hat{\lambda}_{2} \in \mathbb{C}_{-} \cup \mathbb{R}$
are all taken to be distinct, i.e.,
$\lam{1}{} \neq \lam{2}{}$ and $\hat{\lambda}_{1} \neq \hat{\lambda}_{2}$. 
\noindent
\textbf{{\large Case I:}}
\newline
If all eigenvalues in the complex plane are pure imaginary, that is
$\lam{1}{}=\i m_{1}$,
$\lam{2}{}=\i m_{2}$,
$\hat{\lambda}_{1}=-\i m_{1}$, 
$\hat{\lambda}_{2}=-\i m_{2}$, for
$m_{1},m_{2}>0$ 
and
$\wvec{1}{}=(\wvec{11}{},\wvec{12}{},-\wvec{12}{},\wvec{12}{},-\wvec{12}{})^{T}$, 
then for simplicity of the solution, we take
$\wvec{2}{}=\wvec{1}{}$.
Hence, the solution in this nonlocal reverse-spacetime case is given by:
\begin{align}\label{u2solitons-SASA}
\u{}{}(x,t)&= \pm \i \alpha (m_{1}^{2}-m_{2}^{2})
\frac{\Nsol{1}{}(x,t)}{\Dsol{1}{}(x,t)},
\end{align}
where 
\begin{flalign}\nonumber 
\Nsol{1}{}(x,t) &=
m_{2} 
e^{(\alpha_{1}+\alpha_{2}) m_{2} x 
- (\beta_{1}+\beta_{2}) m_{2}^{5} t}
\Big(
e^{-2 (\alpha_{1} m_{1} x +\beta_{1} m_{1}^{5} t)}
+
e^{-2 (\alpha_{2} m_{1} x +\beta_{2} m_{1}^{5} t)}
\Big)
\\ & 
-m_{1} 
e^{(\alpha_{1}+\alpha_{2}) m_{1} x 
- (\beta_{1}+\beta_{2}) m_{1}^{5} t}
\Big(
e^{-2 (\alpha_{1} m_{2} x +\beta_{1} m_{2}^{5} t)}
+
e^{-2 (\alpha_{2} m_{2} x +\beta_{2} m_{2}^{5} t)}
\Big) 
\\
\nonumber
\Dsol{1}{}(x,t)&=
(m_{1}-m_{2})^{2} 
\Big(
e^{-2 \alpha_{1} (m_{1}+m_{2}) x 
-2 \beta_{1} (m_{1}^{5}+m_{2}^{5}) t}
+
e^{-2 \alpha_{2} (m_{1}+m_{2}) x 
-2 \beta_{2} (m_{1}^{5}+m_{2}^{5}) t}
\Big)
\\ & \nonumber 
+(m_{1}+m_{2})^{2}
\Big(
e^{-2 (\alpha_{1} m_{1} +\alpha_{2} m_{2}) x 
-2 (\beta_{1} m_{1}^{5} +\beta_{2} m_{2}^{5}) t}
+
e^{-2 (\alpha_{1} m_{2} +\alpha_{2} m_{1}) x 
-2 (\beta_{1} m_{2}^{5} +\beta_{2} m_{1}^{5}) t}
\Big)
\\ &
-8 m_{1} m_{2}
e^{- (\alpha_{1}+\alpha_{2}) (m_{1}+m_{2}) x 
- (\beta_{1}+\beta_{2}) (m_{1}^{5}+m_{2}^{5}) t}. &
\end{flalign}
\subsubsection{\textbf{Dynamics of the two-soliton solution: Case I}}
\begin{figure}[H]
\begin{center}
\fcolorbox{gray}{black}{\includegraphics[width=0.45\textwidth,height=4.5cm]{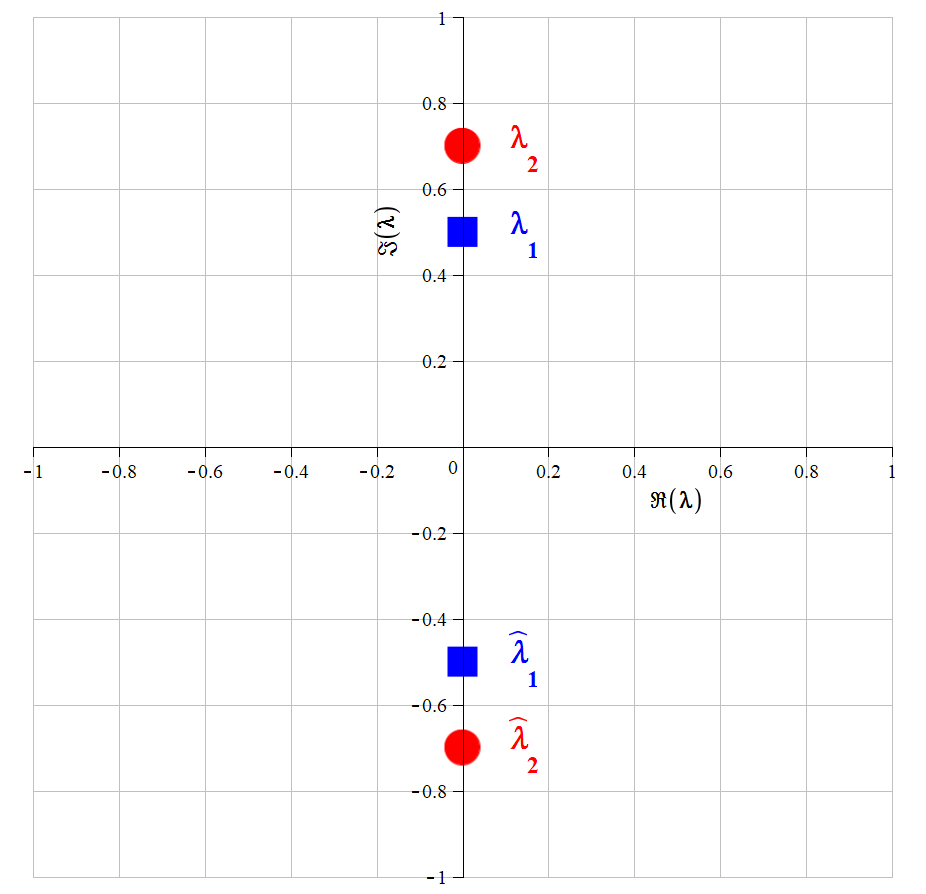}}
\fcolorbox{gray}{black}{\includegraphics[width=0.45\textwidth,height=4.5cm]{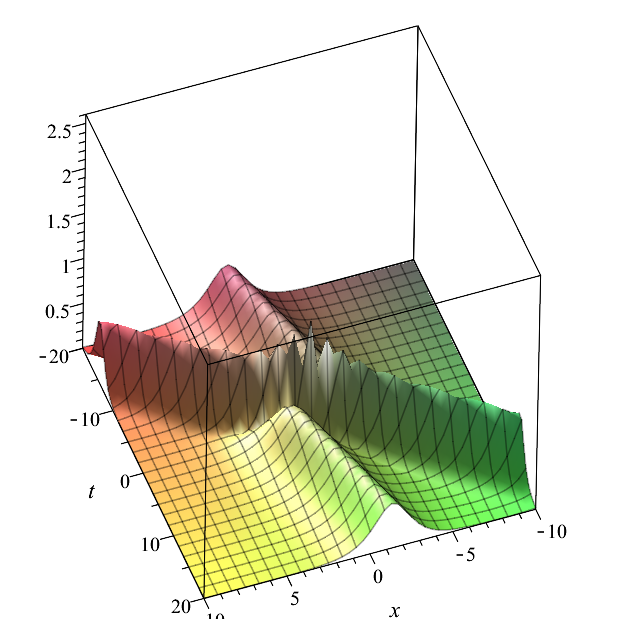}}
\\
\fcolorbox{gray}{black}{\includegraphics[width=0.45\textwidth,height=4.5cm]{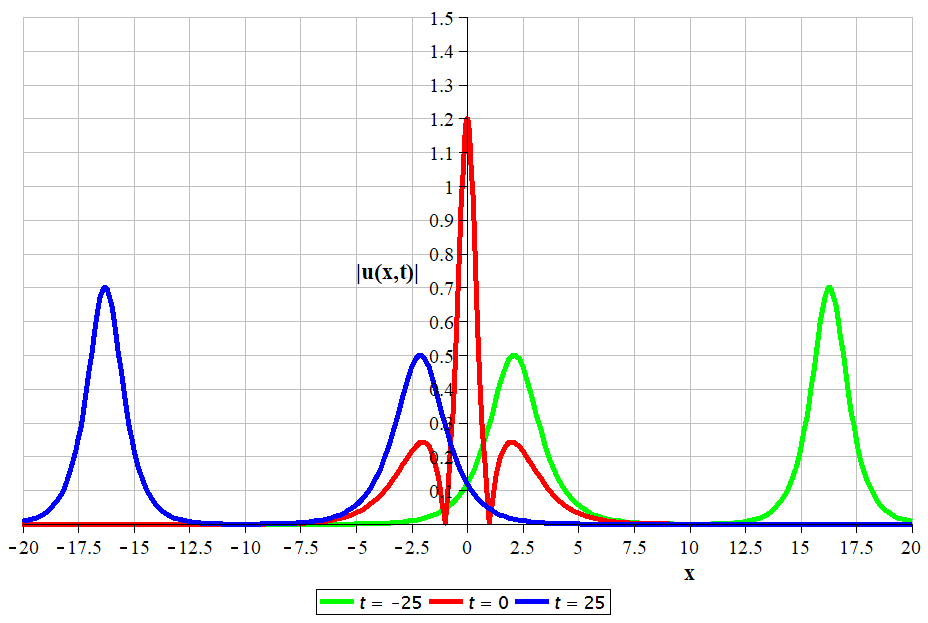}}
\fcolorbox{gray}{black}{\includegraphics[width=0.45\textwidth,height=4.5cm]{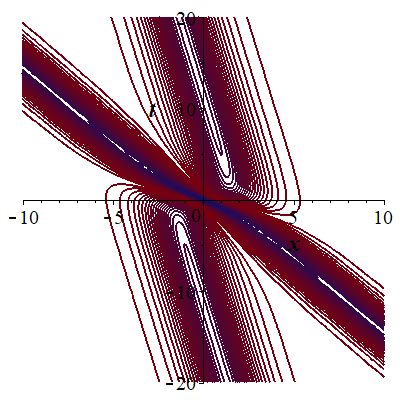}}
\caption{Spectral plane along with 3D, 2D and contours plots of $|u(x,t)|$ of the two solitons interaction with parameters  
$(\alpha,\beta)=(-2,-5)$, 
$(\lam{1}{},\lam{2}{},\lamhat{1}{},\lamhat{2}{})=
(0.5\i,0.7\i,-0.5\i,-0.7\i)$, 
$\wvec{1}{}=\wvec{2}{}=(2,1,-1,1,-1)$.}
\label{2solitonplot1case1-SASA}
\end{center}
\end{figure}

\noindent
If the eigenvalues $\lam{1}{}=-\lamhat{1}{}$ and $\lam{2}{}=-\lamhat{2}{}$,
then the two solitons move in the same direction before and after collision, where the faster soliton overtakes the slower one. 
An overlay of two traveling waves is shown in figure \ref{2solitonplot1case1-SASA}, where the faster soliton overtakes the slower one from the right. Furthermore, in the pre and post collision the amplitude remains unchanged.
\noindent
\textbf{{\large Case II:}}
\newline
In that case, if $\lam{1}{},\lam{2}{} \in \mathbb{C}_{+}$
are not pure imaginary, then the involution property (\ref{lambdacondition-SASA}) requires that $\lam{2}{}=-\lambar{1}{}$,
while in the lower half-plane 
$\hat{\lambda}_{1}=- \lam{1}{}$
and
$\hat{\lambda}_{2}= \lambar{1}{}$.
\newline
Let
$\wvec{1}{}=(\wvec{11}{},\wvec{12}{},\wvec{12}{},\wvec{12}{},\wvec{12}{})^{T}$
and
$\wvec{2}{} = \wvec{1}{}$.
the solution in this nonlocal reverse-spacetime case
is given by: 
\begin{align}\label{u2solitonscase2-SASA}
\u{}{}(x,t)&= 
\i 4 \alpha \imag(\lam{}{2}) \wvec{11}{} \wvec{12}{}
\frac{\Nsol{2}{}(x,t)}{\Dsol{2}{}(x,t)},
\end{align}
where 
\begin{flalign} \nonumber
\Nsol{2}{}(x,t)&=
2 \wvec{11}{2} \real \Big( \lambar{}{} 
e^{\i \big(
2 \alpha_{1} \lam{}{} - (\alpha_{1}+\alpha_{2}) \lambar{}{} 
\big) x
+ \i \big(
2 \beta_{1} \lam{}{5} - (\beta_{1}+\beta_{2}) \lambar{}{5}
\big) t}
\Big)
\\ & \hspace{4cm}
+8 \wvec{12}{2} \real \Big( \lambar{}{} 
e^{\i \big(
2 \alpha_{2} \lam{}{} - (\alpha_{1}+\alpha_{2}) \lambar{}{} 
\big) x
+ \i \big(
2 \beta_{2} \lam{}{5} - (\beta_{1}+\beta_{2}) \lambar{}{5}
\big) t}
\Big).
&
\end{flalign}
\newpage
\begin{flalign} \nonumber
\Dsol{2}{}(x,t)&=
4 (\real(\lam{}{}))^{2} \wvec{11}{4}
e^{-4 \alpha_{1} \imag(\lam{}{}) x -4 \beta_{1} \imag(\lam{}{5}) t}
+64 (\real(\lam{}{}))^{2} \wvec{12}{4}
e^{
-4 \alpha_{2} \imag(\lam{}{}) x
-4 \beta_{2} \imag(\lam{}{5}) t }
\\& \nonumber
-32 (\imag(\lam{}{}))^{2} \wvec{11}{2} \wvec{12}{2}
\real \Big( 
e^{\i 2 (\alpha_{2} \lam{}{} -\alpha_{1} \lambar{}{}) x
+\i 2 (\beta_{2} \lam{}{5} -\beta_{1} \lambar{}{5}) t}
\Big)
\\ & 
+32 |\lam{}{}|^{2} \wvec{11}{2} \wvec{12}{2}
e^{
-2 (\alpha_{1}+\alpha_{2}) \imag(\lam{}{}) x
-2 (\beta_{1}+\beta_{2}) \imag(\lam{}{5}) t}. &
\end{flalign}
\subsubsection{\textbf{Dynamics of the two-soliton solution: Case II}}
\begin{figure}[H]
\begin{center}
\fcolorbox{gray}{black}{\includegraphics[width=0.45\textwidth,height=4.5cm]{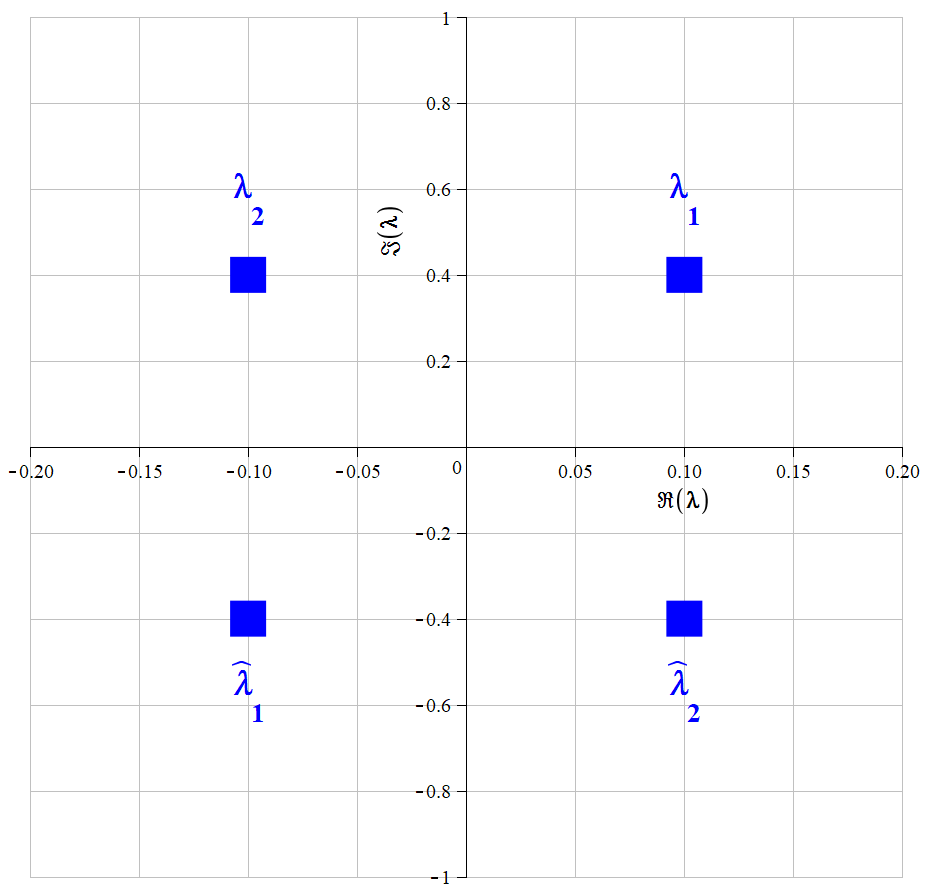}}
\fcolorbox{gray}{black}{\includegraphics[width=0.45\textwidth,height=4.5cm]{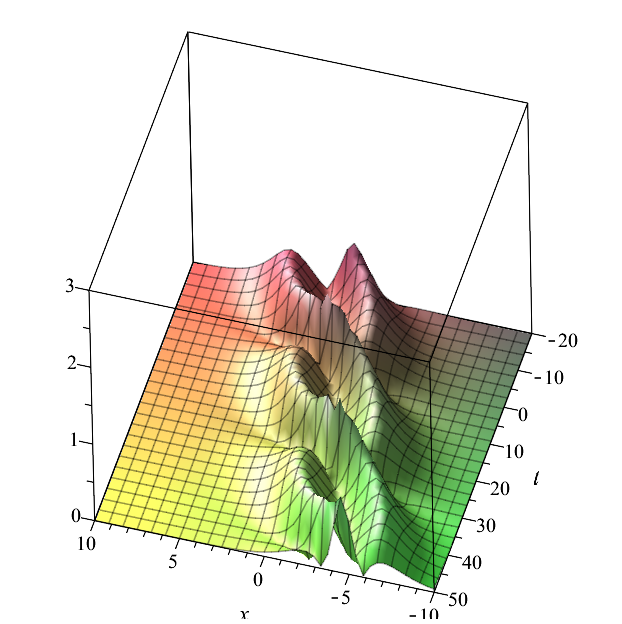}}
\\
\fcolorbox{gray}{black}{\includegraphics[width=0.45\textwidth,height=4.5cm]{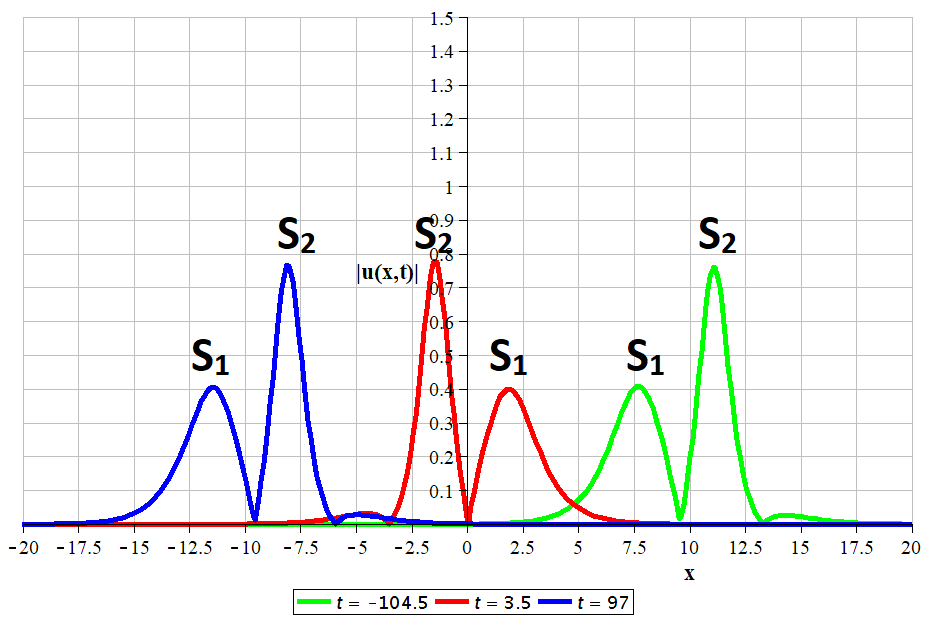}}
\fcolorbox{gray}{black}{\includegraphics[width=0.45\textwidth,height=4.5cm]{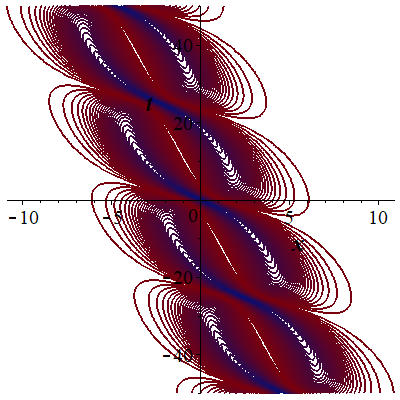}}
\caption{Spectral plane along with 3D, 2D and contours plots
of $|u(x,t)|$ of the two solitons interaction with parameters  $(\alpha,\beta)=(-2,-5)$, 
$(\lam{1}{},\lam{2}{},\lamhat{1}{},\lamhat{2}{})=
(0.1+0.4\i,-0.1+0.4\i,-0.1-0.4\i,0.1-0.4\i)$, 
$\wvec{1}{}=\wvec{2}{}=(2,1,1,1,1)$.}
\label{2solitonplot1case2-SASA}
\end{center}
\end{figure}

\begin{figure}[H]
\begin{center}
\fcolorbox{gray}{black}{\includegraphics[width=0.45\textwidth,height=4.5cm]{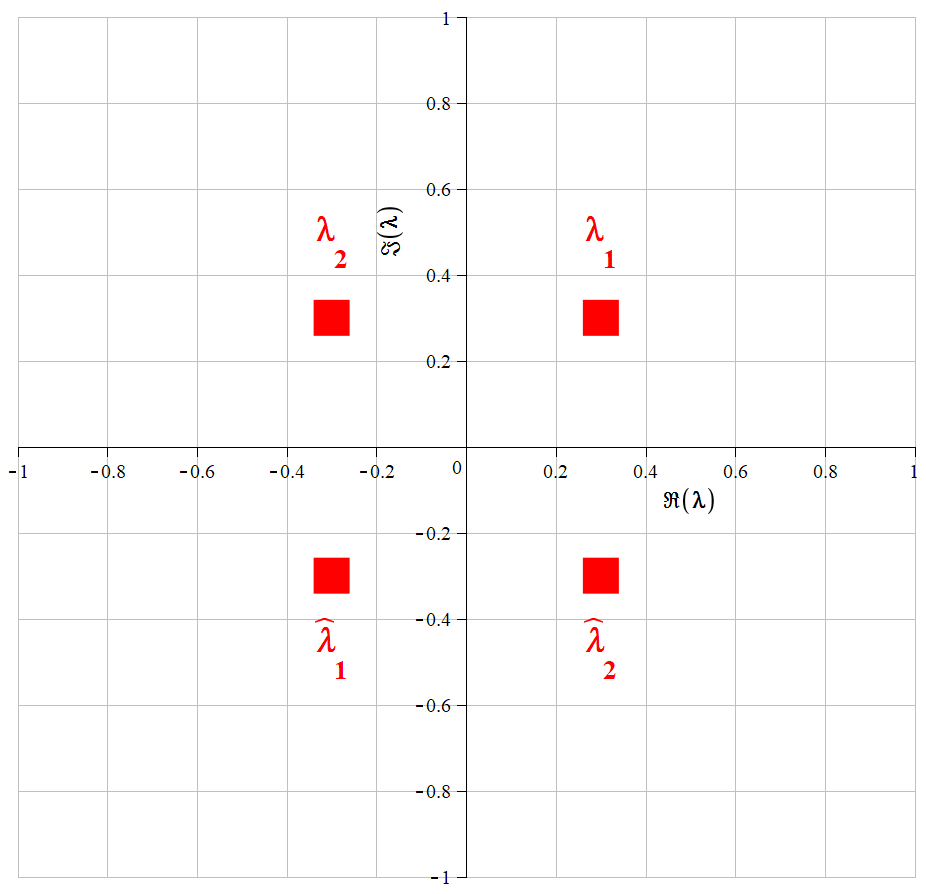}}
\fcolorbox{gray}{black}{\includegraphics[width=0.45\textwidth,height=4.5cm]{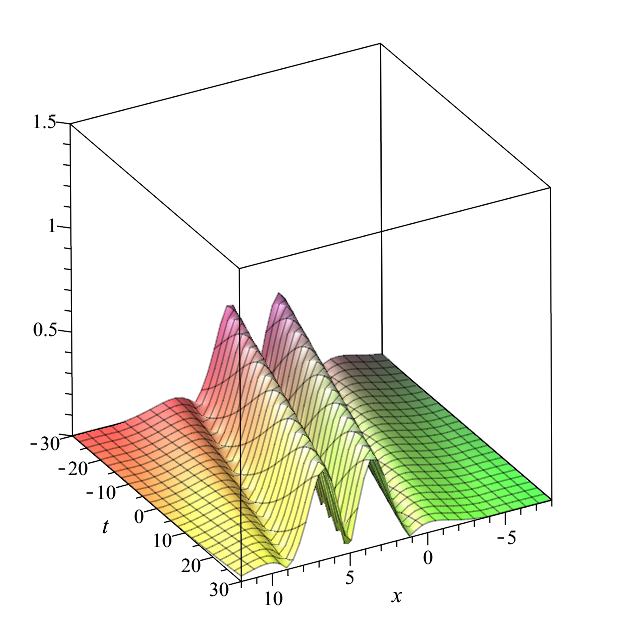}}
\\
\fcolorbox{gray}{black}{\includegraphics[width=0.45\textwidth,height=4.5cm]{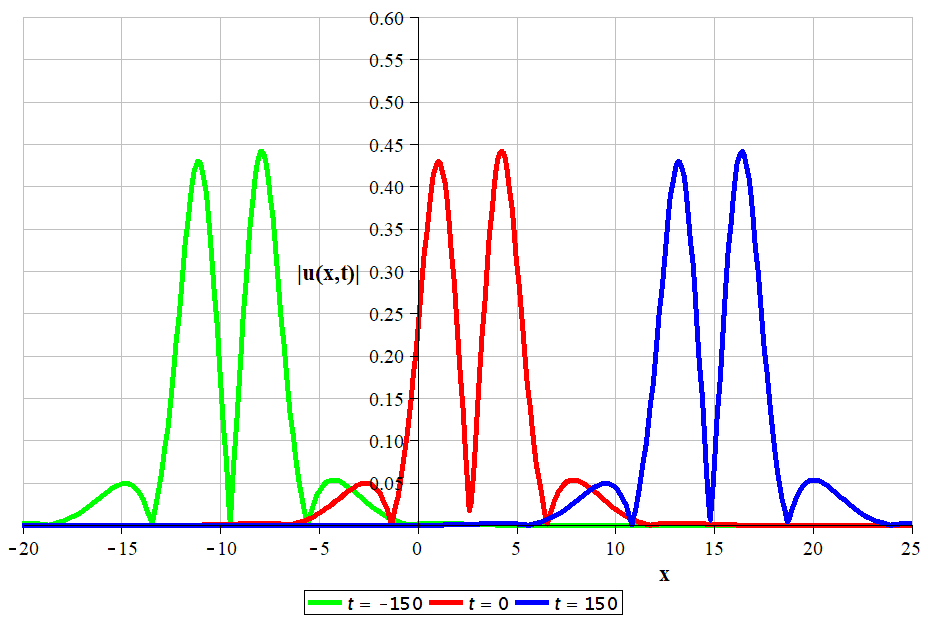}}
\fcolorbox{gray}{black}{\includegraphics[width=0.45\textwidth,height=4.5cm]{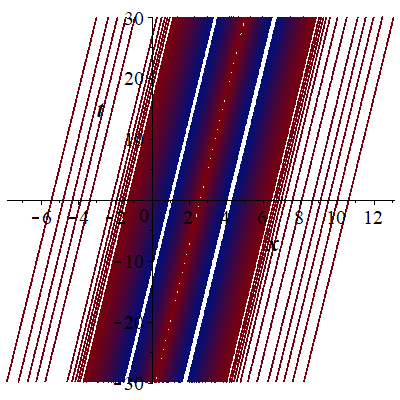}}
\caption{Spectral plane along with 3D, 2D and contours plots
of $|u(x,t)|$ of the two solitons interaction with parameters  $(\alpha,\beta)=(-2,-5)$, 
$(\lam{1}{},\lam{2}{},\lamhat{1}{},\lamhat{2}{})=
(0.1+0.4\i,-0.1+0.4\i,-0.1-0.4\i,0.1-0.4\i)$, 
$\wvec{1}{}=\wvec{2}{}=(2,1,1,1,1)$.}
\label{2solitonplotset3-SASA}
\end{center}
\end{figure}

\noindent
In this configuration of the eigenvalues, the two solitons
$S_{1}$ and $S_{2}$ move in the same direction as shown in figure \ref{2solitonplot1case2-SASA}. 
The soliton wave $S_{2}$ with the higher speed overtakes the wave $S_{1}$ and after the collision, the wave $S_{1}$ gains speed and overtakes $S_{2}$. Therefore, we have a continuously occurring phenomenon of periodic collisions
or an oscillatory-breather.
While in figure \ref{2solitonplotset3-SASA}, we have a two-soliton double-humped keeping the shape as it moves.
\newpage
\noindent
\textbf{{\large Case III:}}
\newline
In that case, if $\lam{1}{} =\i m \in \i \mathbb{R}_{+}$ is pure imaginary and $\lam{2}{} = n \in \mathbb{R}_{+}$,
then the involution property (\ref{lambdacondition-SASA})
requires that $\lamhat{1}{}= - \i m$ and
$\lamhat{2}{}= - n$.
\newline
Let
$\wvec{1}{}=\wvec{2}{}=(\wvec{11}{},\wvec{12}{},-\wvec{12}{},\wvec{12}{},-\wvec{12}{})^{T}$.
The solution for this nonlocal reverse-spacetime case 
reads: 
\begin{align}\label{u2solitonscase3-SASA}
\u{}{}(x,t)&= 
-2 \alpha (m^{2}+n^{2}) \wvec{11}{} \wvec{12}{}
\frac{\Nsol{3}{}(x,t)}{\Dsol{3}{}(x,t)},
\end{align}
where 
\begin{flalign} \nonumber
\Nsol{3}{}(x,t)&=
\i m 
e^{-(\alpha_{1}+\alpha_{2}) m x
-(\beta_{1}+\beta_{2}) m^{5} t}
\Big( \wvec{11}{2}
e^{\i 2 \alpha_{1} n x +\i 2 \beta_{1} n^{5} t}
+ 4 \wvec{12}{2}
e^{\i 2 \alpha_{2} n x +\i 2 \beta_{2} n^{5} t}
\Big)
\\& 
- n 
e^{\i (\alpha_{1}+\alpha_{2}) n x
+ i (\beta_{1}+\beta_{2}) n^{5} t}
\Big( \wvec{11}{2}
e^{-2 \alpha_{1} m x - 2 \beta_{1} m^{5} t}
+ 4 \wvec{12}{2}
e^{-2 \alpha_{2} m x - 2 \beta_{2} m^{5} t}
\Big), &
\end{flalign}
\vspace{-1.5cm}
\begin{flalign} \nonumber
\Dsol{3}{}(x,t)&=
4 (\i m+n)^{2} \wvec{11}{2} \wvec{12}{2}
\Big(
e^{ 2(\i \alpha_{1} n -\alpha_{2} m) x +
2 (\i \beta_{1} n^{5} -\beta_{2} m^{5}) t}
+
e^{ 2(\i \alpha_{2} n -\alpha_{1} m) x +
2 (\i \beta_{2} n^{5} -\beta_{1} m^{5}) t}
\Big)
\\ \nonumber
&
- (\i m +n)^{2} \wvec{11}{4} 
e^{2 \alpha_{1} (\i n -m) x
+2 \beta_{1} (\i n^{5} -m^{5}) t}
- 16 (\i m +n)^{2} \wvec{12}{4} 
e^{2 \alpha_{2} (\i n -m) x
+2 \beta_{2} (\i n^{5} -m^{5}) t}
\\&
-\i 32 mn \wvec{11}{2} \wvec{12}{2}
e^{(\alpha_{1}+\alpha_{2}) (\i n -m) x 
+(\beta_{1}+\beta_{2}) (\i n^{5} -m^{5}) t}. &
\end{flalign}
\newpage
\subsubsection{\textbf{Dynamics of the two-soliton solution: Case III}}
\begin{figure}[H]
\begin{center}
\fcolorbox{gray}{black}{\includegraphics[width=0.45\textwidth,height=4.5cm]{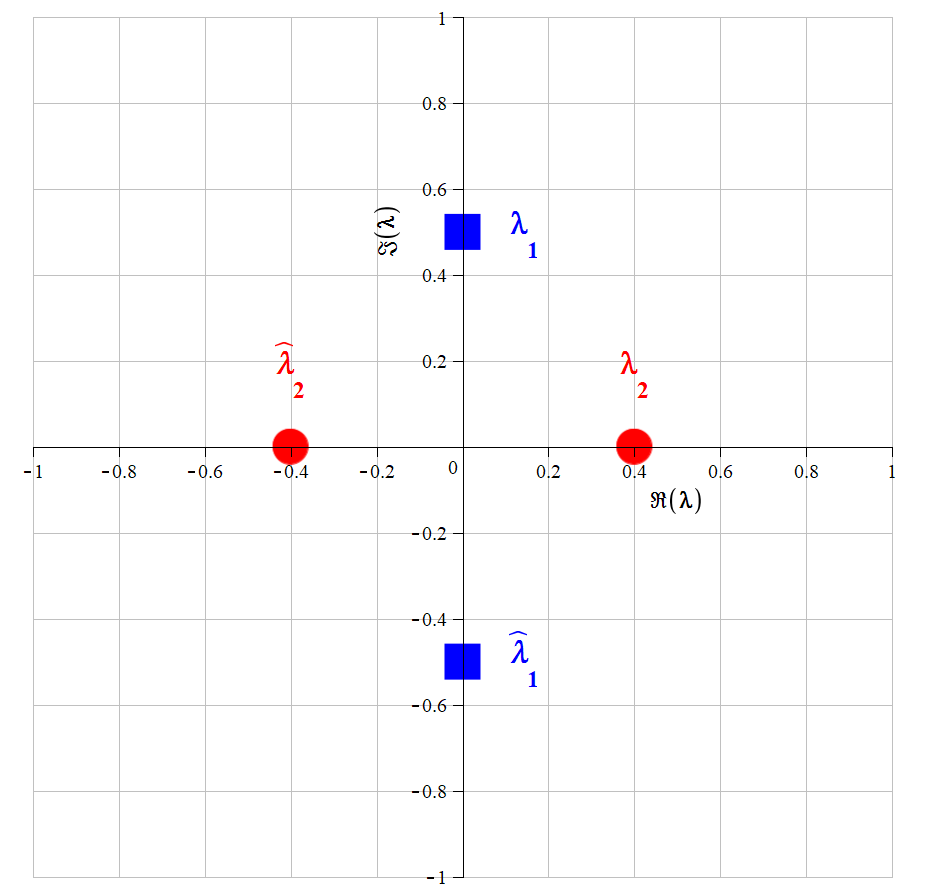}}
\fcolorbox{gray}{black}{\includegraphics[width=0.45\textwidth,height=4.5cm]{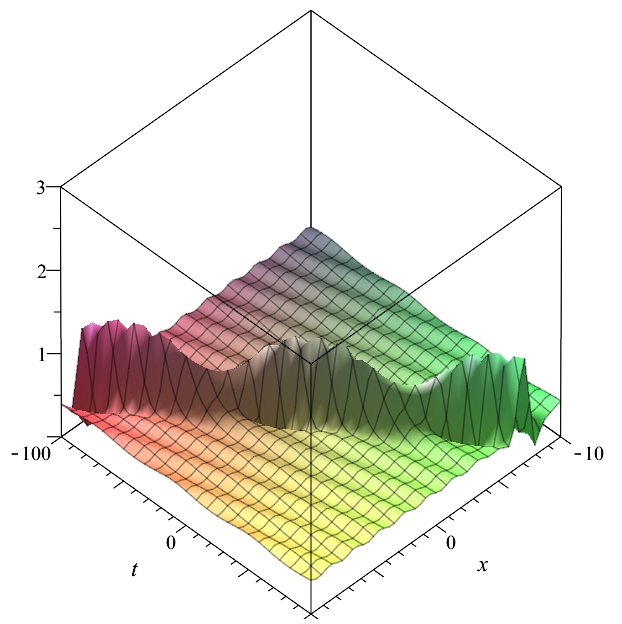}}
\\
\fcolorbox{gray}{black}{\includegraphics[width=0.45\textwidth,height=4.5cm]{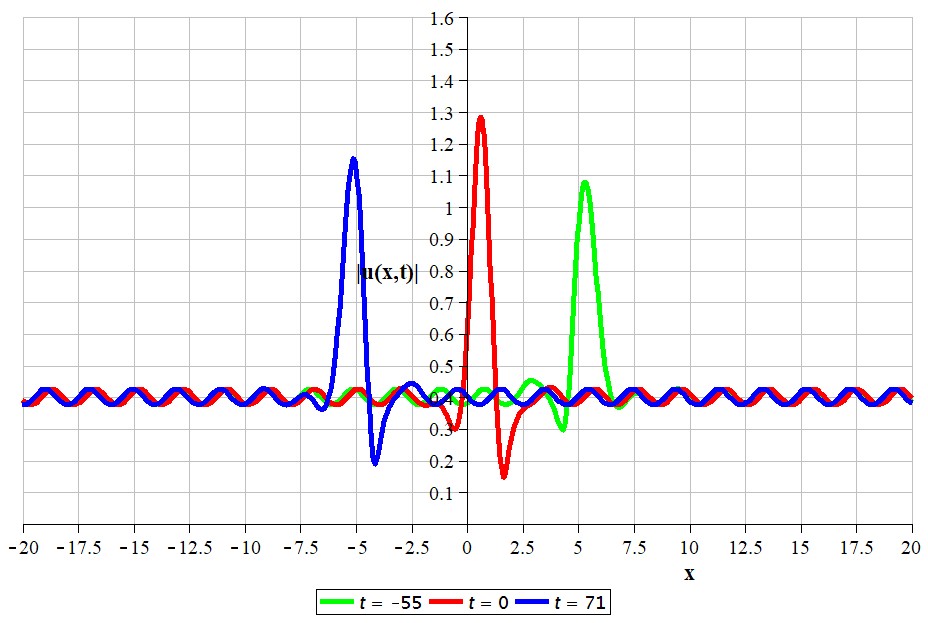}}
\fcolorbox{gray}{black}{\includegraphics[width=0.45\textwidth,height=4.5cm]{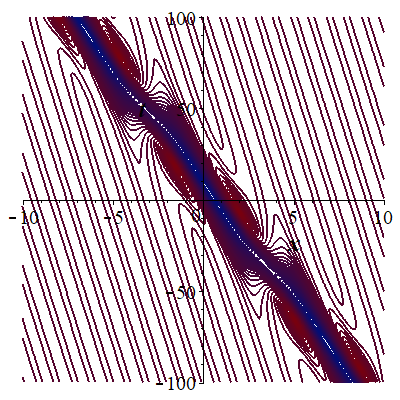}}
\caption{Spectral plane along with 3D, 2D and contours plots
of $|u(x,t)|$ of the continuous interaction between the soliton wave and the breather. The parameters are
$(\alpha, \beta)=(-4,-5)$, 
$(\lam{1}{},\lam{2}{},\lamhat{1}{},\lamhat{2}{})=
(0.5 \i,0.4,-0.5 \i,-0.4)$, 
$\wvec{1}{}=\wvec{2}{}=(2,4,-4,4,-4)$.}
\label{2solitonset7breather-SASA}
\end{center}
\end{figure}

\begin{figure}[H]
\begin{center}
\fcolorbox{gray}{black}{\includegraphics[width=0.45\textwidth,height=4.5cm]{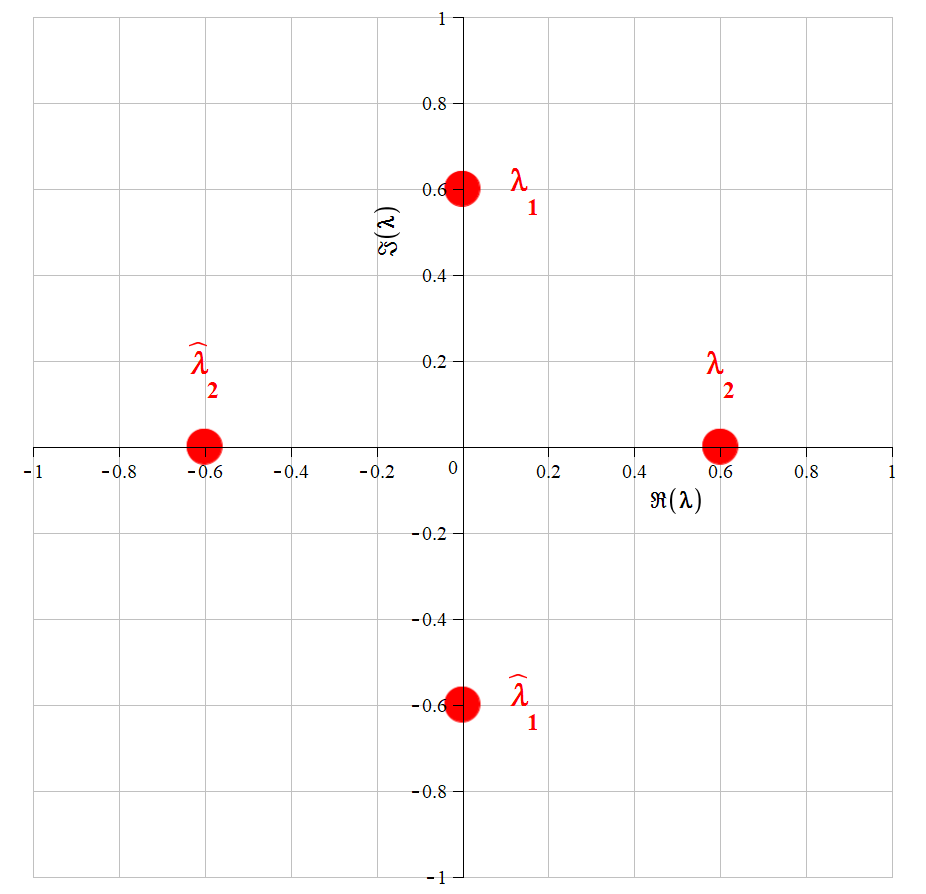}}
\fcolorbox{gray}{black}{\includegraphics[width=0.45\textwidth,height=4.5cm]{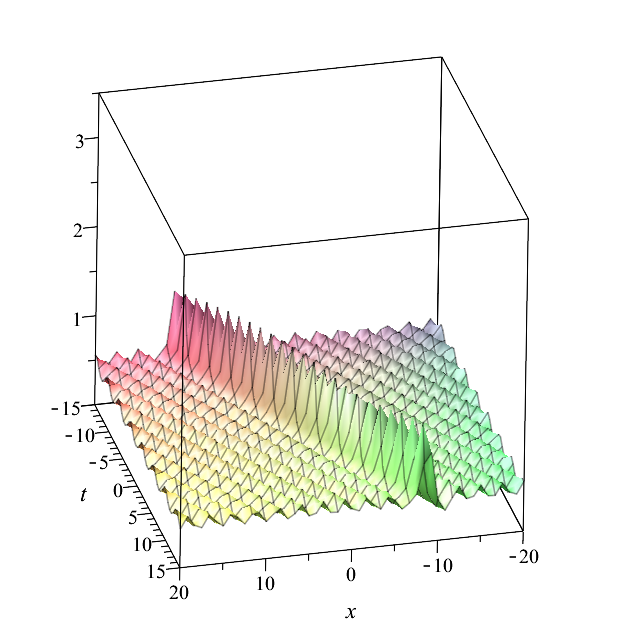}}
\\
\fcolorbox{gray}{black}{\includegraphics[width=0.45\textwidth,height=4.5cm]{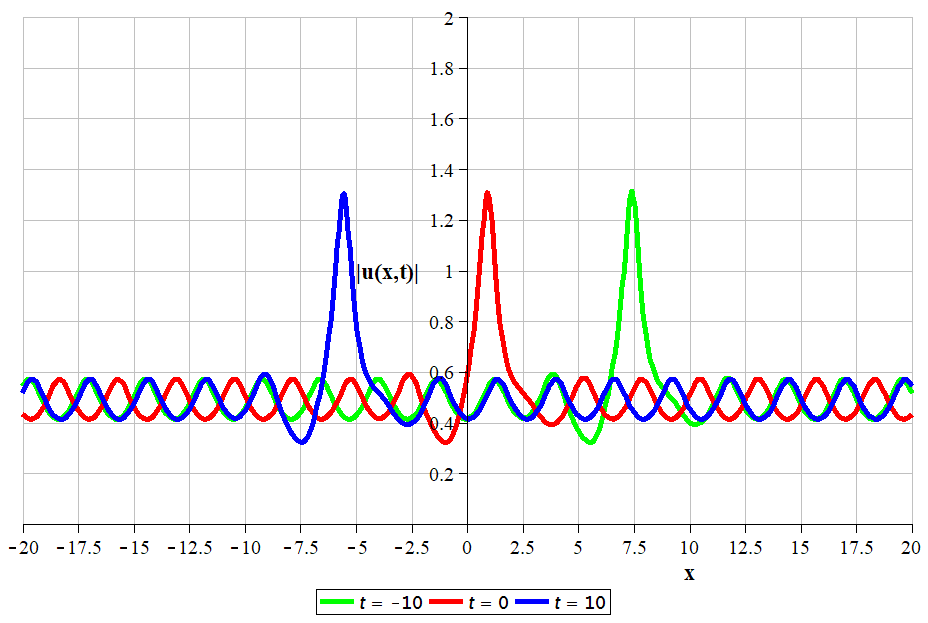}}
\fcolorbox{gray}{black}{\includegraphics[width=0.45\textwidth,height=4.5cm]{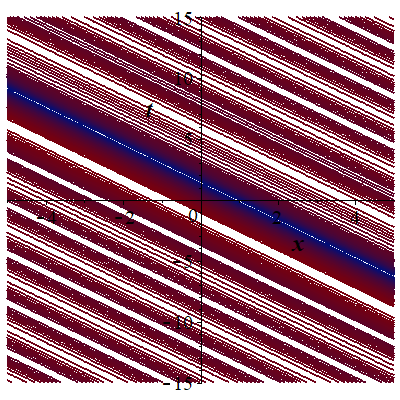}}
\caption{Spectral plane along with 3D, 2D and contours plots
of $|u(x,t)|$ of the soliton wave and the breather. The parameters are
$(\alpha, \beta)=(-2,-10)$, 
$(\lam{1}{},\lam{2}{},\lamhat{1}{},\lamhat{2}{})=
(0.6 \i,0.6,-0.6 \i,-0.6)$, 
$\wvec{1}{}=\wvec{2}{}=(2,2.5,-2.5,2.5,-2.5)$.}
\label{2solitonplot1breather-SASA}
\end{center}
\end{figure}

\noindent
Taking a look at this dynamics, we can observe a soliton
and a breather moving in the same direction.
They interact continuously while the soliton travels through the breather. This is shown in figure \ref{2solitonset7breather-SASA}. 
Another possible different dynamics is when a soliton and a breather travel together at the same speed without interacting as seen in figure \ref{2solitonplot1breather-SASA}, that is the soliton
in the latter situation keeps its shape at all time while moving with the breather as a packet.
\newpage
\noindent
\textbf{{\large Two soliton breather:}}
\newline
In this particular case, the configuration (\ref{EigenvaluesConfiguration-SASA}) compels a two-soliton breather to behave as a one-soliton breather, if all eigenvalues are real. That is, since
$\lam{1}{} \neq \lam{2}{}$ and $\hat{\lambda}_{1} \neq \hat{\lambda}_{2}$,
then $\lam{2}{}$ and $\lamhat{2}{}$ are redundant and we take 
$\lam{2}{}=\lamhat{2}{}=0$, which reduces to the one-soliton breather solution, previously mentioned (figure \ref{1solitonplot4-SASA}).
\begin{figure}[H]
\centering
\includegraphics[width=5cm,height=5cm]{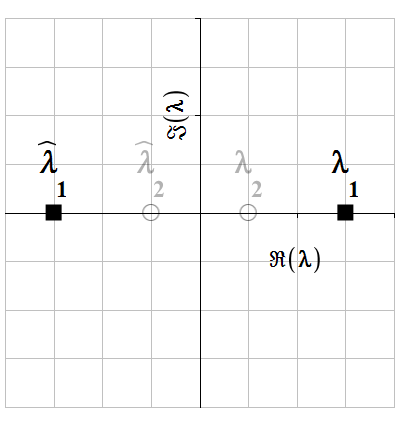}
\caption{Spectral plane of the two-soliton breather}
\label{2-soliton-Sasa-Satsuma-case4-SASA}
\end{figure}
\newpage
\section{Conclusion}
To summarize, in this paper, we investigated a nonlocal reverse-spacetime two-component Sasa-Satsuma equation. This equation is derived from a nonlocal integrable hierarchy, where the nonlocal nature is embodied within the hierarchy's structure. The latter construction allows nonlocal systems to be constructed without using reductions and guarantees integrability. 
Moreover, the hierarchy provide mKdV-type nonlocal integrable
equations and eliminate NLS-type ones.
Futher, a kind of soliton solutions was generated, and the Hamiltonian structure was derived for the resutling nonlocal Sasa-Satsuma equation. 
\\
For fundamental soliton solutions to local equations, solitons interact elastically in a superposition manner, while for soliton solutions to nonlocal equations, this is not always the case. Also, for nonlocal equations, soliton solutions can have singularities at a finite time, but not for the presented Sasa-Satsuma euqtion. 
\newline
Moreover, reverse-spacetime equations exhibit very different dynamical behaviours
than reverse-time and reverse-space equations \cite{Yang2019}. For instance, it can be seen from the plotted
figures that the one-soliton to the reverse-spacetime Sasa-Satsuma equation is a moving soliton, while there is a stationary one-soliton in the reverse-time and reverse-space NLS equation \cite{AlleAhmedMa}.
\newline
Finally, we remark that it remains intriguing to solve nonlocal integrable equations in the cases of reverse-spacetime, reversespace, and reverse-time by different techniques such as Darboux transformations and the Hirota bilinear method \cite{MatveevSalle1991}-\cite{SunMaYu}.


\end{document}